\documentclass[useAMS,usenatbib]{mnras}
\usepackage{graphicx,subfig}
\usepackage{float}
\usepackage{bm}
\usepackage{amsmath}
\usepackage{amsfonts,amssymb}
\usepackage{color}
\definecolor{v}{rgb}{0.6, 0.2, 0.8} 
\usepackage{soul} 
\usepackage{enumerate}
\usepackage{float}
\usepackage{subfig}
\usepackage{tabularx}
\usepackage{multicol}        
\usepackage{bm}	
\usepackage{hyperref}

\title[The Cardassian expansion revisited: constraints from OHD and SN Ia data]{The Cardassian expansion revisited: constraints from updated Hubble parameter measurements and Type Ia Supernovae data}
\author[Maga\~na, H. Amante, Garc\'{\i}a-Aspeitia and Motta]{Juan Maga\~na$^{1}$ \thanks{E-mail:juan.magana@uv.cl}, Mario H. Amante$^{2,1}$\thanks{E-mail:mario.herrera@fisica.uaz.edu.mx}, Miguel A. Garcia-Aspeitia$^{3,2}$\thanks{E-mail:aspeitia@fisica.uaz.edu.mx}, V. Motta$^{1}$\thanks{E-mail: veronica.motta@uv.cl} \\
$^{1}$Instituto de F\'isica y Astronom\'ia, Facultad de Ciencias, \\Universidad de Valpara\'iso, Avda. Gran Breta\~na 1111, Valpara\'iso, Chile. \\
$^{2}$ Unidad Acad\'emica de F\'isica, Universidad Aut\'onoma de Zacatecas, \\Calzada Solidaridad esquina con Paseo a la Bufa S/N C.P. 98060,Zacatecas, M\'exico.\\
$^{3}$Consejo Nacional de Ciencia y Tecnolog\'ia, Av. Insurgentes Sur 1582. \\ Colonia Cr\'edito Constructor, Del. Benito Ju\'arez C.P. 03940, Ciudad de M\'exico, M\'exico.}

\date{Accepted YYYYMMDD. Received YYYYMMDD; in original form YYYYMMDD}
\pubyear{2017}

\begin{document}

\maketitle

\begin{abstract}
Motivated by an updated compilation of observational Hubble data (OHD) which consist of 51 points in the redshift range $0.07<z<2.36$, we study an interesting model known as Cardassian which drives the late cosmic acceleration without a dark energy component. Our compilation contains 31 data points measured with the differential age method by \citet{Jimenez:2001gg}, and 20 data points obtained from clustering of galaxies.
We focus on two modified Friedmann equations: the original Cardassian (OC) expansion and the modified polytropic Cardassian (MPC). The dimensionless Hubble, $E(z)$, and the deceleration parameter, $q(z)$, are revisited in order to constrain the OC and MPC free parameters, first with the OHD and then contrasted with recent observations of SN Ia using the compressed and full joint-light-analysis (JLA) samples. We also perform a joint analysis using the combination OHD plus compressed JLA. Our results show that the OC and MPC models are in agreement with the standard cosmology and naturally introduce a cosmological-constant-like extra term in the canonical Friedmann equation with the capability of accelerating the Universe without dark energy.
\end{abstract}


\section{Introduction}

The cold dark matter with a cosmological constant ($\Lambda$CDM) model is the cornerstone of modern cosmology. It has shown an unprecedented success predicting and reproducing 
the dynamics and evolution of the Universe. 
$\Lambda$CDM is based on two important but unknown components, dark matter (DM) and dark energy (DE), 
which constitute $\sim96\%$ of the total content of our Universe \citep{Planck2015XIII:2016}. 
In this standard paradigm, the DE, responsible of the late cosmic acceleration, is supplied 
by a cosmological constant (CC), which is associated to vacuum energy. Although several cosmological observations favor the CC, some theoretical problems arise when we try a renormalization of the quantum vacuum fluctuations using an appropriate cut-off at the Planck energy. However, the problem becomes insurmountable, giving a difference of $\sim120$ orders in magnitude between theory and observations \citep{Weinberg,Zeldovich}. In addition, the problem of coincidence, i.e. the similitude between the energy density of matter and dark energy at the present epoch, remains as an open question \citep{Weinberg,Zeldovich}.

To overcome these problems, several alternatives to the CC are proposed, such as quintessence, phantom energy, Chaplygin gas, holographic DE, Galileons, among others \citep[see][for a complete review]{Copeland:2006wr,Carroll:2000fy}. Geometrical approaches are also used to explain the DE dynamics (i.e. brane theories) like Dvali, Gabadaze and Porrati \citep[DGP,][]{Deff:2002}, Randall-Sundrum I and II \citep[RSI,RSII,][]{RaSuI:1999,RaSuII:1999} or $f(R)$ theories \citep{Buchdahl:1970,STAROBINSKY198099,Cembranos:2008gj}; 
each one having important pros and cons.

An interesting alternative, closely related to geometrical models, is the Cardassian expansion model for which there is no DE and the late cosmic acceleration is driven by the modification of the Friedmann equation as $H^2=f(\rho)$ \citep{Xu:2012ig}, where $f(\rho)$ is a functional form of the energy density of the Universe. \citet{Freese:2002} proposed $f(\rho)\propto\rho+\rho^{n}$ in order to obtain a late acceleration stage under certain conditions on the $n$ parameter, naming the model as the Cardassian expansion \footnote{The name Cardassian refers to a humanoid race in Star Trek series, whose goal is the accelerated expansion of their evil empire. This race looks foreign to us and yet is made entirely of matter \citep{Freese:2002}} (hereafter the original Cardassian, OC, model). However,  this expression can be naturally deduced from extra dimensional theories (DGP, RSI, RSII, etc.), which imprint the effects of a 5D space-time (the bulk) in our 4D space-time (the brane) at cosmological scales. In the case of the DGP model, a consequence of this kind of geometry 
is a density parameter that evolves as $(\sqrt{\rho+\alpha}+\beta)^2$, where $\alpha$ and $\beta$ are constants related to the threshold between the brane and the bulk, allowing an accelerated epoch driven only by geometry. In the case of RS models, 
a quadratic term in the energy momentum tensor
modifies the right-hand-side of the Friedmann equation as $a\rho+b\rho^2$ \citep{sms}, with a correspondence to the Cardassian models when $n=2$. 
Thus, the topological structure of the brane and the bulk can naturally produce the Cardassian Friedmann equation. Indeed, it is possible to obtain a $n$-energy-momentum tensor from a Gauss equation with a product of $n$-extrinsic curvatures, which leads to the $\rho^n$ extra term in the Friedmann equation of the original Cardassian model. Therefore, the model motivation is based on extra dimensions arising from a fundamental theory \citep[for an excellent review of extra dimensions models, see for instance][or \citealt{m2000} for a cosmological point of view]{Maartens:2003tw}. Another alternative interpretation is to consider a fluid (that may or may not be in an intrinsically four-dimensional metric) with an extra contribution to the energy-momentum tensor \citep{Gondolo:2003}. Both interpretations are interesting and the standard cosmological dynamics can be mimicked without the need to postulate a dark energy component . In addition, we notice that it is possible to recover a CC when $\rho^n\to1$, without adding it by hand. An OC model generalization
can be obtained by considering an additional exponent in the right-hand-side of the Friedmann equation as $f(\rho)\propto\rho(1+\rho^{l(n-1)})^{1/l}$ which is called modified polytropic Cardassian (hereafter MPC) model by analogy with a fluid interpretation \citep{Gondolo:2002}.

The Cardassian models are extensively studied in the literature. They have been tested with several cosmological observations \citep[see for example][and references therein]{Wang:2003,Wei:2015kua,Liang:2011,Feng:2009zf,Xu:2012ig,Li:2012ApJ}. \citet{Wei:2015kua} put constraints on the OC model parameters using a joint analysis of gamma ray burst data and Type Ia supernovae (SN Ia) of the Union 2.1 sample \citep{Suzuki:2012}. Recently, \citet{Magana:2015A} used the strong lensing measurements in the galaxy cluster Abell 1689, baryon acoustic oscillations (BAO), cosmic microwave background (CMB) data from nine-year Wilkinson microwave anisotropy probe (WMAP) observations \citep{Hinshaw:2013}, and the SN Ia LOSS sample \citep{Ganeshalingam:2013mia} to constrain the MPC parameters.  

In this work, we revisit the Cardassian expansion models with
an universe that contains baryons, DM, together with the radiation component. We explore two functional forms of the Friedmann equation: one with the OC parameter $n$ \citep[following][]{Freese:2002}, and the other one considering also the $l$ exponent \citep[following][]{Gondolo:2003}.
These Cardassian models are tested using an update sample of observational Hubble parameter data (OHD) and the compressed joint-light-analysis (cJLA) SN Ia data by \citet{Betoule:2014}.

As a final comment, while we were finalizing this paper,
an arxiv submitted article \citep[Ref.][]{Zhai:2017} addressed a similar revision of the Cardassian models. While the main focus of Ref. \citet{Zhai:2017} is to match the
the Cardassian Friedmann equations to $f(T,\mathcal{T})$ theory through the action principle, our work focus on providing bounds to the Cardassian models using OHD \citep[see also][]{Zhaiz:2017}. Nonetheless, the authors also provide constraints derived from SN Ia, BAO, and CMB data.

The paper is organized as follows. In Sec. \ref{EM} the Cardassian cosmology is revisited, introducing two proposals for the Friedmann equation, which correspond to the OC and MPC models, and the deceleration parameter is calculated.  In Sec. \ref{data}, we present the data and methodology in order to study the Cardassian models using OHD and SN Ia observations. In Sec. \ref{Res}, we show the constraints for the free parameters 
presenting the novel contrast with the updated sample. Finally,   Sec. \ref{Con} presents our conclusions and the possible outlooks into future studies.

We will henceforth use units in which $c=\hbar=1$ (unless explicitly written).

\section{The Cardassian Cosmology} \label{EM}

\subsection{Original Cardassian model}

The original Cardassian model was introduced by \citet{Freese:2002}
to explain the accelerated expansion of the Universe without DE.
Motivated by braneworld theory, this model modifies the Friedmann
equation as 
\begin{equation}
H^{2}=\frac{8\pi G \rho_{t}}{3} + B\rho_{t}^{n}, \label{C1}
\end{equation}
where $H=\dot{a}/a$ is the Hubble parameter, $a$ is the scale factor of the Universe, $G$ is the Newtonian gravitational constant, $B$ is a dimensional coupling constant that depends on the theory, and the total matter density is $\rho_{t}=\rho_{m}+\rho_{r}$. The recent Planck measurements \citep{Ade:2013zuv,Planck2015XIII:2016}  suggest
a curvature energy density $\Omega_k\simeq 0$, thus we assume a flat geometry.
The conservation equation is maintained in the traditional form:
\begin{equation}
\dot{\rho}+3H(\rho+p)=0. \label{CE}
\end{equation}
 The matter density (dark matter and baryons), $\rho_{m}=\rho_{m0}a^{-3}$,
and the radiation density, $\rho_{r}=\rho_{r0}a^{-4}$, evolution can be computed from Eq. \eqref{CE}.
The second term in the right hand side of Eq. (\ref{C1}), known as the Cardassian term, 
drives the universe to an accelerated phase if the exponent $n$ satisfies $n<2/3$. At early times, this corrective term is negligible and the dynamics of the universe is governed by the canonical term of the Friedmann equation. When the universe evolves, the traditional energy density and the one due to the Cardassian correction becomes equal at redshift $z_{Card}\sim \mathcal{O}(1)$. Later on, the 
Cardassian term begins to dominate the evolution of the universe and source the cosmic acceleration. 
The Eq. (\ref{C1}) reproduces the $\Lambda$CDM model for $n= 0$.
As in the standard case, it is possible to define a new critical density for the OC model, $\rho_{Nc}$, which satisfies the Eq. (\ref{C1}) and can be written as $\rho_{Nc}=\rho_{c}F(B,n)$, where $\rho_{c}=3H^{2}/8\pi G$ is the standard critical density, and $F(B,n)$ is a function which depends on the OC parameters and the components of the Universe.

The Raychaudhuri equation can be written in the form:
\begin{equation}
\frac{\ddot{a}}{a}=-\frac{4\pi G}{3}(\rho_t+3p_t)-B\left[\left(\frac{3n}{2}-1\right)\rho_t^n+\frac{3}{2}n\rho_t^{n-1}p_t\right],
\end{equation}
where Eqs. \eqref{C1} and \eqref{CE} were used. From Eq. \eqref{C1}, it is possible to obtain the dimensionless Hubble parameter $E^2(z)\equiv H^2(z)/H^2_0$ as, 
\begin{equation}
E(z,\mathbf{\Theta})^{2}= \Omega_{\mathrm{std}}+(1-\Omega_{m0}-\Omega_{r0})\left[\frac{\Omega_{\mathrm{std}}}{\Omega_{m0}+\Omega_{r0}}\right]^{n},
\label{eq:EzCar}
\end{equation}
\noindent
where $\mathbf{\Theta}=(\Omega_{m0},h,n)$ is the free parameter vector to be constrained by the data, $\Omega_{r0}=\rho_{r0}/\rho_{c}$ is the current standard density parameter for the radiation component, $\Omega_{m0}=\rho_{m0}/\rho_{c}$ is the observed standard density parameter for matter (baryons and DM), and we define $\Omega_{\mathrm{std}}\equiv\Omega_{m0}(1+z)^{3} + \Omega_{r0}(1+z)^{4}$. We compute $\Omega_{r0}=2.469\times10^{-5}h^{-2}(1+0.2271 N_{eff})$ \citep{Komatsu:2011}, where $N_{eff}=3.04$ is the standard number of relativistic species \citep{Mangano:2002}. Notice that we have also imposed a flatness condition on the total content of the Universe \citep[for further details on how to obtain Eq. (\ref{eq:EzCar}) see][]{SenA:2003,SenS:2003}.

The deceleration parameter, defined as $q\equiv-\ddot{a}/aH^2$, can be written as
\begin{eqnarray}
&&q(z,\mathbf{\Theta})=\frac{\Omega_{\mathrm{std}}-\frac{1}{2}\Omega_{m0}(1+z)^3}{E^2(z,\mathbf{\Theta})}+\frac{1-\Omega_{m0}-\Omega_{r0}}{(\Omega_{m0}+\Omega_{r0})^n}\times \nonumber\\
&&\left[\left(\frac{3n}{2}-1\right)+\frac{n\Omega_{r0}}{2\Omega_{std}}(1+z)^4\right]\frac{\Omega_{\mathrm{std}}^n}{E^2(z,\mathbf{\Theta})}.
\end{eqnarray}

In order to investigate whether the OC model can drive the late cosmic acceleration, it is necessary to reconstruct the $q(z)$ using the mean values for the $\mathbf{\Theta}$ parameters. 

\subsection{Modified polytropic Cardassian model}

\citet{Gondolo:2002,Gondolo:2003} introduced a simple generalization of the Cardassian
model, the modified polytropic Cardassian, by introducing an
additional exponent $l$ \citep[see also][]{Wang:2003}. The modified
Friedmann equation with this generalization can be written as
\begin{equation}
H^{2} = \frac{8\pi G}{3} \rho_{t} \beta^{1/l},
\label{eq:Hmpc}
\end{equation}
where
\begin{equation}
\beta\equiv 1 + \left( \frac{\rho_{Card}}{\rho_{t}} \right)^{l(1-n)},
\end{equation}
and $\rho_{Card}$ is the characteristic energy density, with $n<2/3$ and $l>0$. In concordance with the previous Friedmann Eq. \eqref{C1} and following \cite{Ade:2013zuv,Planck2015XIII:2016}, we also assume $\Omega_k\simeq 0$.
The Eq. (\ref{eq:Hmpc}) reproduce the $\Lambda$CDM model for $l = 1$ and $n= 0$. Thus, the acceleration equation is
\begin{equation}
\frac{\ddot{a}}{a}=-\frac{4\pi G}{3}\rho_{t}\beta^{1/l}+4\pi G(1-n)\rho_{t}\left(1-\frac{1}{\beta}\right)\beta^{1/l}. \label{ac2}
\end{equation}
The MPC model (Eq. \ref{eq:EzMPC}) has been studied by several authors using different data with $\rho_{t}=\rho_{m}$ \citep[see for example][]{Feng:2009zf} and also with $\rho_{t}=\rho_{m}+\rho_{r}$ together with a curvature term \citep{Shi:2012}. Here we consider a flat MPC with matter and radiation components.
After straightforward calculations, the dimensionless $E^{2}(z,\mathbf{\Theta})$ parameter reads as:
\begin{eqnarray}
&&E^{2}(z,\mathbf{\Theta})=\Omega_{r0}(1+z)^{4} + \Omega_{m0}(1+z)^{3}\beta(z,\mathbf{\Theta})^{1/l},\quad
\label{eq:EzMPC}
\end{eqnarray}
where
\begin{equation}
\beta(z,\mathbf{\Theta})\equiv1 + \left[ \left( \frac{1-\Omega_{r0}}{\Omega_{m0}}\right)^l  - 1 \right](1+z)^{3l(n - 1)},
\end{equation}
being $\mathbf{\Theta}=(\Omega_{m0},h,l,n)$, the free parameter vector to be fitted by the data. 

In addition, $q(z,\mathbf{\Theta})$, can be written as
\begin{eqnarray}
&&q(z,\mathbf{\Theta})=\frac{\Omega_{m0}\beta(z,\mathbf{\Theta})^{1/l}}{2E^2(z,\mathbf{\Theta})}\left[1-3(1-n)\left(1-\frac{1}{\beta(z,\mathbf{\Theta})}\right)\right]\times\nonumber\\&&(1+z)^3+\frac{\Omega_{r0}}{E^2(z,\mathbf{\Theta})}(1+z)^4.
\end{eqnarray}
We use the $\mathbf{\Theta}$ mean values in the last expression to reconstruct
the deceleration parameter $q(z)$ and investigate whether the MPC model is consistent with a late cosmic acceleration. 

\section{Data and Methodology} \label{data}
The OC and MPC model parameters are constrained using 
an updated OHD sample, which contains $51$ data points, and the compressed SN Ia data set from the JLA full sample by \citet{Betoule:2014}, which contains $31$ data  points. In the following we briefly introduce these
data sets. 

\subsection{Observational Hubble data}\label{subsec:hz}

The ``differential age" (DA) method proposed by \cite{Jimenez:2001gg} allows us to measure the expansion rate of the Universe at redshift $z$, i.e.  $H(z)$. This technique compares the ages of early-type-galaxies (i.e., without ongoing star formation) with similar metallicity and separated by a small redshift interval \citep[for instance,][measure $\Delta z \sim 0.04$ at $z < 0.4$ and $\Delta z \sim 0.3$ at $z>0.4$]{Moresco:2012jh}. Thus, a $H(z)$ point can be estimated using 
\begin{equation}
H(z)=-\frac{1}{1+z}\frac{\mathrm{dz}}{\mathrm{dt}},
\end{equation}
where $\mathrm{dz/dt}$ is measured using the $4000 \mbox{\AA}$ break ($D4000$) feature as function of redshift. A strong $D4000$ break  depends on the metallicity and the age of the stellar population of the early-type galaxy. Thus, the technique by \citet{Jimenez:2001gg} offers to directly measure the Hubble parameter using spectroscopic dating of passively-evolving galaxy to compare their ages and metallicities, providing  $H(z)$ measurements that are model-independent. These H(z) points are given by different authors as
\citet{Zhang:2012mp,Moresco:2012jh,Moresco:2015cya,Moresco:2016mzx,Stern:2009ep}, and constitute the majority of our sample (31 points). In addition, we use 20 points from BAO measurements, although some of them being correlated because they either belong to the same analysis or there is overlapping among the galaxy samples, through this work, we
assume that they are independent measurements. Moreover, some data points are biased because they are estimated using a sound horizon, $r_{d}$\footnote{the sound horizon is the maximum comoving distance which sound waves could travel at redshift $z_{d}$}, at the drag epoch, $z_{d}$, which depends on the cosmological model \citep{Melia:2017}. Points provided by different authors use different values for the $r_{d}$ in clustering measurements, for instance \citet{Anderson:2013oza} take $153.19$ Mpc while \citet{Gaztanaga:2008xz} choose $153.3$ Mpc, etc. 

Table \ref{tab:Hz} shows an updated compilation of OHD accumulating a total of 51 points \citep[other recent compilations are provided by][]{Farooq:2016zwm, Zhang:2016tto,Yu:2016gmd}. We have included all the points of the previous references, although priority has been given to the measurements that comes from the DA method and have also been measured with clustering at the same redshift. As reference to compare our results, we also give the data point by \citet{Riess:2016jrr} who measured a Hubble constant $H_0$ with $2.4\%$ of uncertainty. Authors argue that this improvement is due to a better calibration (using Cepheids) of the distance to 11 SN Ia host galaxies, reducing the error by almost one percent. 
We use this sample to constrain the free parameters of the OC and MPC models and look for an alternative solution to the accelerated expansion of the Universe. The figure-of-merit for the OHD is given by

\begin{equation}
\chi_{\mbox{OHD}}^2 = \sum_{i=1}^{N_{OHD}} \frac{ \left[ H(z_{i}) -H_{obs}(z_{i})\right]^2 }{ \sigma_{H_i}^{2} },
\end{equation}
where $N_{OHD}$ is the number of the observational Hubble parameter $H_{obs}(z_{i})$ at $z_{i}$,
$\sigma_{H_i}$ is its error, and $H(z_{i})$ is the theoretical value for a given  model.

\begin{table*}
\begin{tabular}{|lllll|}
\hline
$z$ & $H(z)$ & $\sigma_{H}$ & Reference & Method \\
 & $\mbox{km\,s}^{-1}\mbox{Mpc}^{-1}$ & \tiny{kms$^{-1}$Mpc$^{-1}$} & & \\
\hline
0 & 73.24 & 1.74 & \cite{Riess:2016jrr} & SN Ia/Cepheid \\
\hline
0.07 & 69 & 19.6 & \cite{Zhang:2012mp} & DA\\
0.1 & 69 & 12 & \cite{Stern:2009ep} & DA\\
0.12 & 68.6 & 26.2 & \cite{Zhang:2012mp} & DA\\
0.17 & 83 & 8 & \cite{Stern:2009ep} & DA \\
0.1791 & 75 & 4 & \cite{Moresco:2012jh} & DA\\
0.1993 & 75 & 5 & \cite{Moresco:2012jh} & DA\\
0.2& 72.9 & 29.6 & \cite{Zhang:2012mp} & DA\\
0.24 & 79.69 & 2.65 & \cite{Gaztanaga:2008xz} & Clustering \\
0.27 & 77 & 14 & \cite{Stern:2009ep} & DA\\
0.28 & 88.8 & 36.6 & \cite{Zhang:2012mp} & DA\\
0.3 & 81.7 & 6.22 & \cite{Oka:2013cba}& Clustering \\
0.31 & 78.17 & 4.74 & \cite{Wang:2016wjr}& Clustering \\
0.35 & 82.7 & 8.4 & \cite{Chuang:2012qt} & Clustering\\
0.3519 & 83 & 14 & \cite{Moresco:2012jh} & DA\\
0.36 & 79.93  & 3.39 & \cite{Wang:2016wjr}& Clustering \\
0.38 & 81.5  & 1.9 & \cite{Alam:2016hwk}& Clustering \\
0.3802 & 83  & 13.5  & \cite{Moresco:2016mzx} & DA \\
0.4 & 95 & 17 & \cite{Stern:2009ep} & DA\\
0.4004 & 77 & 10.2 & \cite{Moresco:2016mzx} & DA \\
0.4247 & 87.1 & 11.2 & \cite{Moresco:2016mzx} & DA\\
0.43 & 86.45 & 3.68 & \cite{Gaztanaga:2008xz} & Clustering\\ 
0.44 & 82.6 & 7.8 & \cite{Blake:2012pj} & Clustering\\
0.4497 & 92.8 & 12.9 & \cite{Moresco:2016mzx} &DA \\
0.47 & 89 & 34 & \cite{Ratsimbazafy:2017vga} & DA\\
0.4783 & 80.9 & 9 & \cite{Moresco:2016mzx} &DA \\
0.48 & 97 & 62 & \cite{Stern:2009ep} & DA\\
0.51 & 90.4  & 1.9 & \cite{Alam:2016hwk}& Clustering \\
0.52 & 94.35  & 2.65 & \cite{Wang:2016wjr}& Clustering \\
0.56 & 93.33  & 2.32 & \cite{Wang:2016wjr}& Clustering \\
0.57 & 92.9 & 7.8 & \cite{Anderson:2013oza} & Clustering\\
0.59 & 98.48  & 3.19 & \cite{Wang:2016wjr}& Clustering \\
0.5929 & 104 & 13 & \cite{Moresco:2012jh} & DA\\
0.6 & 87.9 & 6.1 & \cite{Blake:2012pj} & Clustering \\
0.61 & 97.3  & 2.1 & \cite{Alam:2016hwk}& Clustering \\
0.64 & 98.82  & 2.99 & \cite{Wang:2016wjr}& Clustering \\
0.6797 & 92 & 8 & \cite{Moresco:2012jh} & DA\\
0.73 & 97.3 & 7 & \cite{Blake:2012pj} & Clustering\\
0.7812 & 105 & 12 & \cite{Moresco:2012jh} & DA\\
0.8754 & 125 & 17 & \cite{Moresco:2012jh} & DA\\
0.88 & 90 & 40 & \cite{Stern:2009ep} & DA\\
0.9 & 117 & 23 & \cite{Stern:2009ep} & DA\\
1.037 & 154 & 20 & \cite{Moresco:2012jh} & DA\\
1.3 & 168 & 17 & \cite{Stern:2009ep} & DA\\
1.363 & 160 & 33.6 & \cite{Moresco:2015cya} & DA\\
1.43 & 177 & 18 & \cite{Stern:2009ep} & DA\\
1.53 & 140 & 14 & \cite{Stern:2009ep} & DA\\
1.75 & 202 & 40 & \cite{Stern:2009ep} & DA\\
1.965 & 186.5 & 50.4 & \cite{Moresco:2015cya} & DA \\
2.33 & 224 & 8 & \cite{Bautista:2017zgn} & Clustering\\
2.34 & 222 & 7 & \cite{Delubac:2014aqe}& Clustering\\
2.36 & 226 & 8 & \cite{Font-Ribera:2013wce}& Clustering\\ 
\hline
\end{tabular}
\caption{52 Hubble parameter measurements $H(z)$ (in km s$^{-1}$Mpc$^{-1}$) and their errors, $\sigma_{H}$, 
at redshift $z$. The first point is not included in the MCMC analysis, it was only considered as a Gaussian prior in some tests. The method column refers as to how to $H(z)$ was obtained: DA stands for differential age method, and clustering comes from BAO measurements.}
\label{tab:Hz}
\end{table*}

\subsubsection{An homogeneous OHD sample}
\label{subsubsec:homOHD}

As mentioned above, the OHD from clustering (BAO features) are biased
due to an underlying $\Lambda$CDM cosmology to estimate $r_{d}$. 
Different authors used different values in the cosmological parameters and obtained different sound horizons at the drag epoch, which are used to break the degeneracy in $Hr_{d}$. Furthermore, the determination of $H(z)$ from BAO features is computed taking into account very conservative systematic errors \citep[see the discussion by][]{Melia:2017,Meliab:2017}.

As a first attempt to homogenize and achieve model independence for the OHD obtained from clustering, we take the value $Hr_d$ for each data point and assume a common value $r_{d}$ for the entire data set. We consider two $r_{d}$ estimations: $r_{\mathrm{dpl}}= 147.33 \pm 0.49$ Mpc and $r_{\mathrm{dw9}}=152.3 \pm 1.3$ Mpc from the most recent Planck \citep{Planck2015XIII:2016} and WMAP9 \citep{Bennett:2012zja} measurements respectively. 
In addition, we also take into account three other sources of errors that could affect $r_d$ due to its contamination by a cosmological model. The first one comes from the error of each reported value. The second error considers the possible range of $r_{d}$ values provided by separate CMB measurements, i.e. the difference between the sound horizon given by WMAP9 and Planck. This error is the one producing the largest impact on the $r_{d}$ mean value (3.37 \% and 3.26 \% for the Planck and WMAP9 data point respectively). The last error to take into account is the difference between $r_{d}$ used to obtain the OHD and the one that would be obtained if we assume another cosmological model instead of $\Lambda$CDM. Hereafter we use the one obtained for a DE constant equation-of-state ($w$) CDM model, $r_{\mathrm{d\omega cdm}} = 148.38$ Mpc \citep[the cosmological parameters for this model are provided by][]{Neveu:2017}. Adding in quadrature the percentage for these three errors, we obtain $r_{\mathrm{dpl}} = 147.33 \pm 5.08$ Mpc and $r_{\mathrm{dw9}} = 152.3 \pm 6.42$ Mpc. Finally, we propagate this new error to the quantity $H(z)$ to secure a new homogenized and model-independent sample (Table \ref{tab:Hzhom}).

\begin{table}
\centering
\begin{tabular}{|ccc|}
\hline
$z$ & $H(z) \pm \sigma_{H}(r_{\mathrm{dpl}})$ & $H(z)\pm \sigma_{H}(r_{\mathrm{dw9}})$  \\
 & $\mbox{km\,s}^{-1}\mbox{Mpc}^{-1}$ & $\mbox{km\,s}^{-1}\mbox{Mpc}^{-1}$ \\
\hline
0.24 & 82.37 $\pm$ 3.94 & 79.69 $\pm$ 4.28\\
0.3 & 78.83 $\pm$ 6.58 & 76.26 $\pm$ 6.63\\
0.31 & 78.39 $\pm$ 5.46 & 75.83 $\pm$ 5.60\\
0.35 & 88.10 $\pm$ 9.45 & 85.23 $\pm$ 9.37\\
0.36 & 80.16 $\pm$ 4.37 & 77.54 $\pm$ 4.63\\
0.38 & 81.74 $\pm$ 3.40 & 79.08 $\pm$ 3.81\\
0.43 & 89.36 $\pm$ 4.89 & 86.44 $\pm$ 5.18\\
0.44 & 85.48 $\pm$ 8.59 & 82.69 $\pm$ 8.55\\
0.51 & 90.67 $\pm$ 3.66 & 87.71 $\pm$ 4.13\\
0.52 & 94.61 $\pm$ 4.20 & 91.52 $\pm$ 4.63\\
0.56 & 93.59 $\pm$ 3.96 & 90.54 $\pm$ 4.42\\
0.57 & 96.59 $\pm$ 8.76 & 93.44 $\pm$ 8.78\\
0.59 & 98.75 $\pm$ 4.66  & 95.53 $\pm$ 5.07\\
0.6 & 90.96 $\pm$ 7.04 & 87.99 $\pm$ 7.14 \\
0.61 & 97.59 $\pm$ 3.97 & 94.41 $\pm$ 4.47\\
0.64 & 99.09 $\pm$ 4.53 & 95.86 $\pm$ 4.97 \\
0.73 & 100.69 $\pm$ 8.03 & 97.40 $\pm$ 8.12 \\
2.33 & 223.99 $\pm$ 11.12 & 216.69 $\pm$ 11.97 \\
2.34 & 222.105 $\pm$ 10.38 & 214.85 $\pm$ 11.31\\
2.36 & 226.24 $\pm$ 11.18 & 218.86 $\pm$ 12.05\\
\hline
\end{tabular}
\caption{Homogenized model-independent OHD from clustering (in km s$^{-1}$Mpc$^{-1}$) and its error, $\sigma_{H}$, 
              at redshift $z$. The first and second columns were obtained using the sound horizon in the drag epoch from Planck and WMAP measurements respectively.}
\label{tab:Hzhom}
\end{table}
\subsection{Type Ia Supernovae (SN Ia)}\label{subsec:snia}
The SN Ia observations supply the evidence of the accelerated expansion of the Universe. They have been considered a perfect standard candle to measure the geometry and dynamics of the Universe and have been widely used to constrain alternatives cosmological models to explain the late-time cosmic acceleration. Currently, there are several compiled SN Ia samples, for instance, the Union 2.1 compilation by \citet{Suzuki:2012} which consists of 580 points in the redshift range $0.015<z<1.41$, and 
the Lick Observatory Supernova Search (LOSS) sample containing $586$ SN Ia in the redshift range $0.01<z<1.4$ \citep{Ganeshalingam:2013mia}. Recently,  \citet{Betoule:2014} presented the so-called full JLA (fJLA) sample which contains $740$ points spanning a redshift range $0.01<z<1.2$.
The same authors also provide the information of the fJLA data in a compressed set (cJLA) of $31$ binned distance modulus $\mu_{b}$ spanning a redshift range $0.01<z<1.3$, which still remains accurate for some models where the isotropic luminosity distance evolves slightly with redshift. For instance, when the cJLA is used in combination with other cosmological data, the difference between fJLA and cJLA in the mean values for the $w$CDM model parameters is at most $0.018\sigma$. Here we use both, the fJLA and cJLA samples, to constrain the parameters of the OC and MPC models.

\subsubsection{Full JLA sample}
\label{subsubsec:JLAfull}
As mentioned, the full JLA sample contains 740 confirmed SN Ia in the redshift interval $0.01<z<1.2$, which is one
of the most recent and reliable SN Ia samples.  We use this sample to constrain the parameters of both Cardassian models. The function of merit for the fJLA sample is calculated as:
\begin{equation}
\chi^{2}_{\mbox{fJLA}}=\mathbf{\left(\hat{\mu} - \mu_{Card}\right)^{\dag}\mathrm{C_{\eta}^{-1}}\left( \hat{\mu} - \mu_{Card} \right)}, \label{fJLA}
\end{equation}
where  $ \mu_{Card} = 5 log_{10} \left( d_L / 10 pc \right)$, and  $\mathrm{C_{\eta}}$ is the covariance matrix\footnote{available  at  \url{http://supernovae.in2p3.fr/sdss_snls_jla/ReadMe.html}} of $\mathbf{\hat{\mu}}$ provided by  \citet{Betoule:2014}, and is constructed using 
\begin{eqnarray}
\mathbf{C_{\eta}}&&= 
\left( \mathbf{C_{cal}} + \mathbf{C_{model}} + \mathbf{C_{bias}} + \mathbf{C_{host}} + \mathbf{C_{dust}} \right)  \nonumber \\
&& + \left( \mathbf{C_{pecvel}} + \mathbf{C_{nonIa}} \right) + \mathbf{C_{stat}}, \label{CovMat}
\end{eqnarray}
where $\mathbf{C_{cal}}, \mathbf{C_{model}}, \mathbf{C_{bias}}, \mathbf{C_{host}}, \mathbf{C_{dust}}$ are systematic uncertainty matrices associated with the calibration, the light curve model, the bias correction, the mass step, and dust uncertainties respectively. $ \mathbf{C_{pecvel}}$ and $\mathbf{C_{nonIa}}$ corresponds to systematics uncertainties in the peculiar velocity corrections and the contamination of the Hubble diagram by non-Ia events respectively, $\mathbf{C_{stat}}$ corresponds to an statistical uncertainty obtained from error propagation of the light-curve fit uncertainties. Finally $\hat{\mu}$ is given by
\begin{equation}
\hat{\mu} = m_{b}^{\star} - \left( M_{B} -  \alpha \times X_{1}   +  \beta \times C  \right),  \label{muhat}
\end{equation}
where $m_{b}^{\star}$ corresponds to the observed peak magnitude, $\alpha$, $ \beta $ and $M_{B}$ are nuisance parameters in the distance estimates. The $X_{1}$ and $C$ variables describe the time stretching of the light-curve and the Supernova color at maximum brightness respectively. The absolute magnitude $M_{B}$ is related to the host stellar mass ($M_{stellar}$) by the step function:
\begin{align}
M_{B} = \left\{ \begin{array}{cc} 
                M_{B}^{1} & \hspace{5mm}  \rm{if} \ M_{stellar} < 10^{10} M_\odot\ ,\\
              M_{B}^{1} + \Delta_{M} & \hspace{5mm} \rm{otherwise.} \\
                \end{array} \right.
\end{align}
By replacing Eq. (\ref{eq:EzCar}), Eq. (\ref{eq:EzMPC}), Eq. (\ref{CovMat}) and Eq. (\ref{muhat})  in Eq. (\ref{fJLA}), we 
obtain the explicit figure-of-merit $\chi^{2}_{\mathrm{fJLA}}$ for the Cardassian models.\\

\subsubsection{Compressed form of the JLA sample}
\label{subsubsec:cJLA}
Table \ref{tab:jlacom} shows the $31$ binned distance modulus at the binned redshift $z_{b}$.
The function of merit for the cJLA sample is calculated as:
\begin{equation}
\chi^{2}_{\mbox{cJLA}}=\mathbf{r^{\dag}\mathrm{C_{b}^{-1}}r},
\end{equation}
where $\mathrm{C_{b}}$ is the covariance matrix\footnote{available  at  \url{http://supernovae.in2p3.fr/sdss_snls_jla/ReadMe.html}} provided by \citet{Betoule:2014}, and $\mathbf{r}$ is given by
\begin{equation}
\mathbf{r}=\mathbf{\mu_{b}}-M-\log_{10}d_{L}(\mathbf{z_{b}},\mathbf{\Theta}),
\end{equation}
where $M$ is a nuisance parameter and $d_{L}$ is the luminosity distance given by

\begin{equation}
d_{L}=(1+z)\frac{c}{H_{0}}\int^{0}_{z}\frac{\mathrm{dz}^{\prime}}{E(z^{\prime},\mathbf{\Theta})}.
\label{eq:dl}
\end{equation}
By replacing Eq. (\ref{eq:EzCar}) and Eq. (\ref{eq:EzMPC}) in the last expression, we 
obtain the explicit figure of merit $\chi^{2}_{\mathrm{cJLA}}$ for the
OC and MPC models. 

\section{Results} \label{Res}
A MCMC Bayesian statistical analysis was performed to
estimate the ($\Omega_{m0}$,$h$,$n$) and the ($\Omega_{m0}$,$h$,$n$,$l$) parameters for the OC and MPC models respectively.
The constructed Gaussian likelihood function for each data set are given by
$\mathcal{L}_{\mathrm{OHD}}\propto \exp(-\chi_{\mathrm{OHD}}^{2}/2)$, 
$\mathcal{L}_{\mathrm{cJLA}}\propto \exp(-\chi_{\mathrm{cJLA}}^{2}/2)$, $\mathcal{L}_{\mathrm{fJLA}}\propto \exp(-\chi_{\mathrm{fJLA}}^{2}/2)$, and $\mathcal{L}_{\mathrm{joint}}\propto \exp(-\chi_{\mathrm{tot}}^{2}/2)$, where $\chi_{\mathrm{tot}}^{2}=\chi_{\mathrm{OHD}}^{2}+\chi_{\mathrm{cJLA}}^{2}$.
We use the Affine Invariant Markov chain Monte Carlo (MCMC) Ensemble sampler 
from the \emph{emcee} Python module \citep{emcee:2013}. In all our computations we consider $3000$ steps to stabilize the estimations (burn-in phase), $6000$ MCMC steps and $1000$ walkers which are initialized in a small ball around the expected points of maximum probability, is estimated with a differential evolution method. For both, OC and MPC models, we assume the following flat priors: $\Omega_{m0} [0,1]$, and $n [-1,2/3]$. For the $l$ MPC parameter we consider the flat prior $[0,6]$. For the $h$ parameter three priors are considered: a flat prior $[0,1]$, and two Gaussian priors, one by \citet[][the first point in Table \ref{tab:Hz}]{Riess:2016jrr}, and the other one by \citet{Planck2015XIII:2016} from Planck 2015 measurements ($h=0.678\pm0.009$).
When the cJLA data are used, we also take a flat prior on the nuisance parameter $M[-1,1]$. The following flat priors $\alpha[0,2]$, $\beta[0,4.0]$, $M^{1}_{B}[-20,-18]$, and $\Delta_{M}[-0.1,0.1]$ are considered when the fJLA sample is employed. To judge the convergence of the sampler, we ask that the acceptance fraction is in the $[0.2-0.5]$ range and check the autocorrelation time which is found to be $\mathcal{O}(60-80)$.

We carry out four runs using different OHD sets: the full observational sample given in Table \ref{tab:Hz}, the $31$ data points obtained using the DA method ($\mathrm{OHD_{DA}}$), and two samples containing the DA points plus those homogenized points from clustering using a common $r_{d}$ estimated from Planck and WMAP measurements (Table \ref{tab:Hzhom}). We also estimate the OC and MPC parameters using both the cJLA and fJLA samples. Moreover, we perform a joint analysis considering each OHD sample and the cJLA sample. Tables \ref{tab:ocparnueva} and \ref{tab:MPCparnueva} provide the best fits for the OC and MPC parameters respectively using the different data sets and priors on $h$. Tables \ref{tab:ocparjoint} and \ref{tab:mpcparjoint} give the constraints from the following joint analysis: OHD+cJLA (J1), $\mathrm{OHD_{DA}}$+cJLA (J2), $\mathrm{OHD_{hpl}}$+CJLA (J3), and $\mathrm{OHD_{hw9}}$+CJLA (J4). We also give the minimum chi-square, $\chi_{min}$, and the reduced $\chi_{red}=\chi_{min}/d.o.f$, where the degree of freedom (d.o.f.) is the difference between the number of data points and the free parameters. 

\subsection{cJLA vs. fJLA on the Cardassian parameter estimations}
\begin{figure}
\centering
\includegraphics[width=5cm,scale=0.45]{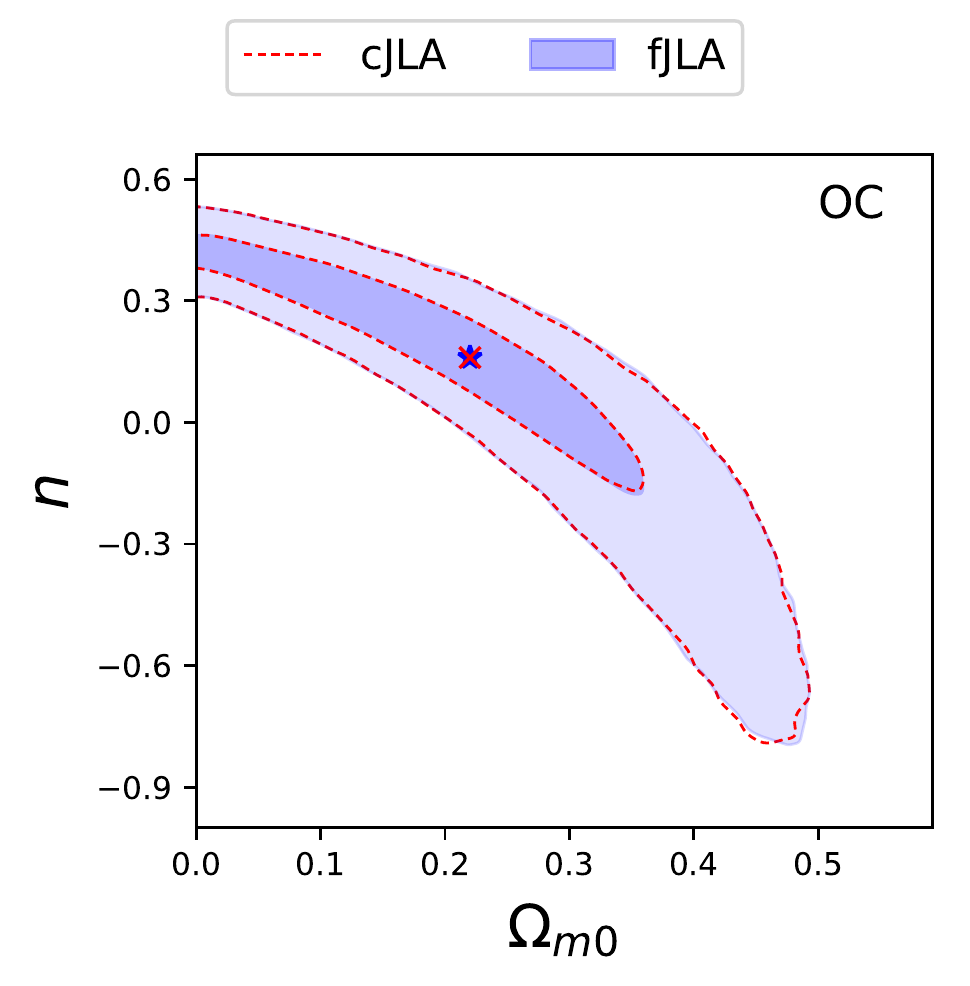} 
\includegraphics[width=5cm,scale=0.45]{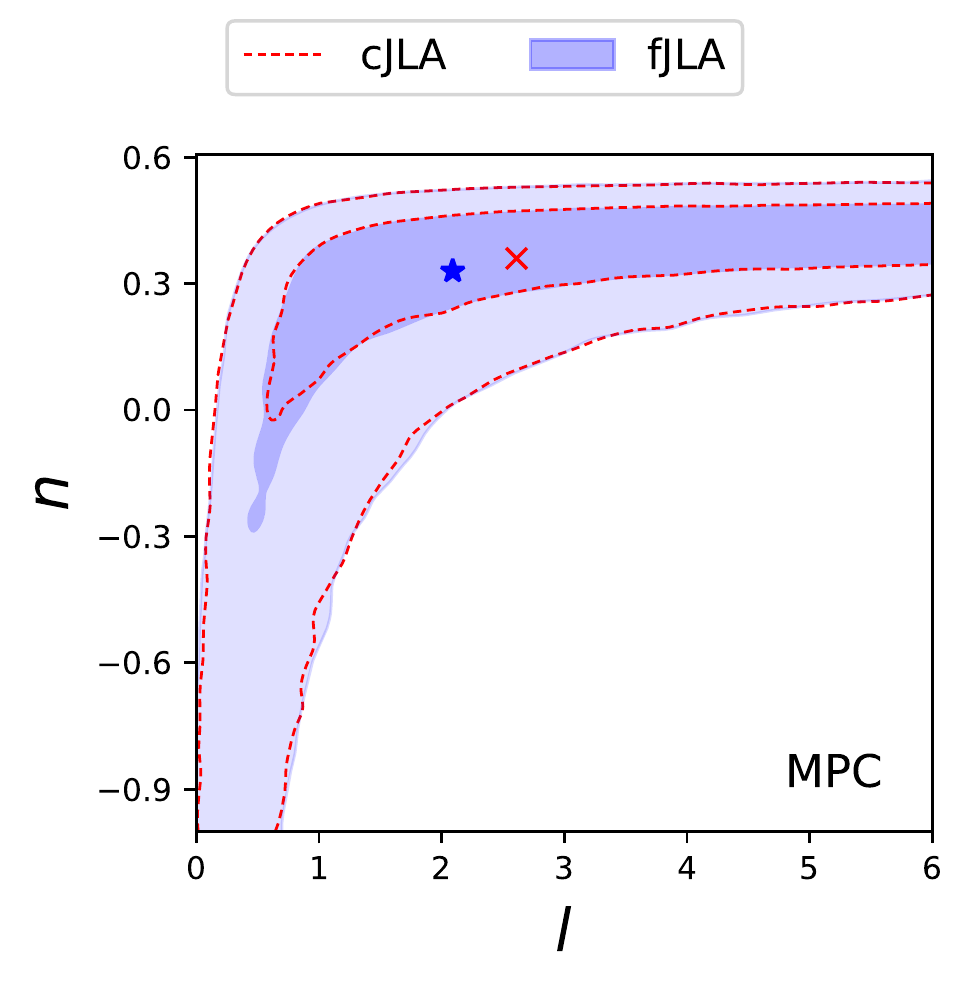} 
\caption{Comparison of the $\Omega_{m0}$-$n$ (top panel) and $n$-$l$ (bottom panel) confidence contours for the OC and MPC parameters within the $1\sigma$ and $3\sigma$ confidence levels using the cJLA (dashed lines) and fJLA (filled and solid lines) samples respectively. In the parameter estimation, a flat prior is considered. The cross and star mark the mean values for each data set.
\label{fig:fjlavscjla}}
\end{figure}

\begin{figure}
  \centering
          \includegraphics[width=6cm,scale=0.45]{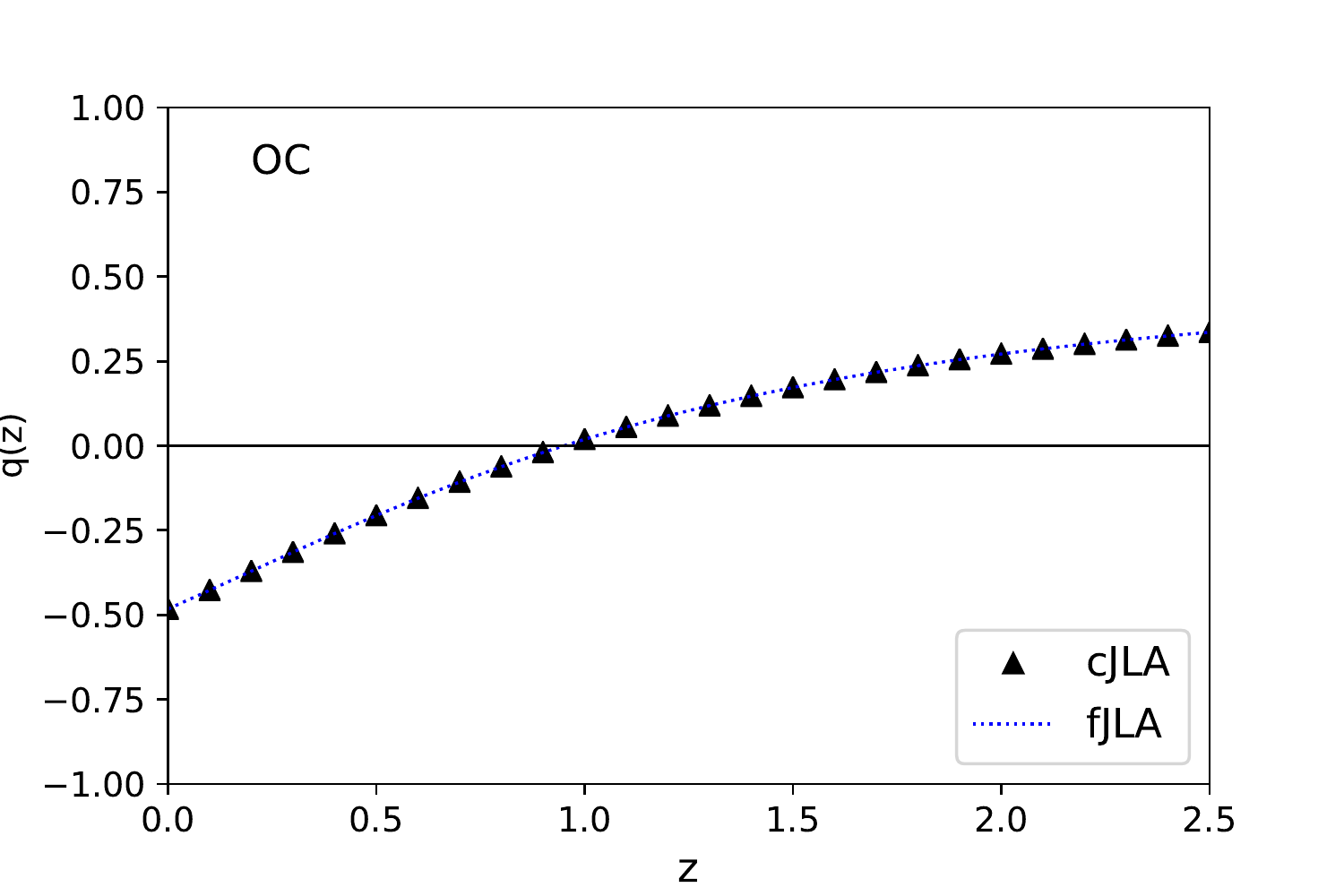} 
            \includegraphics[width=6cm,scale=0.45]{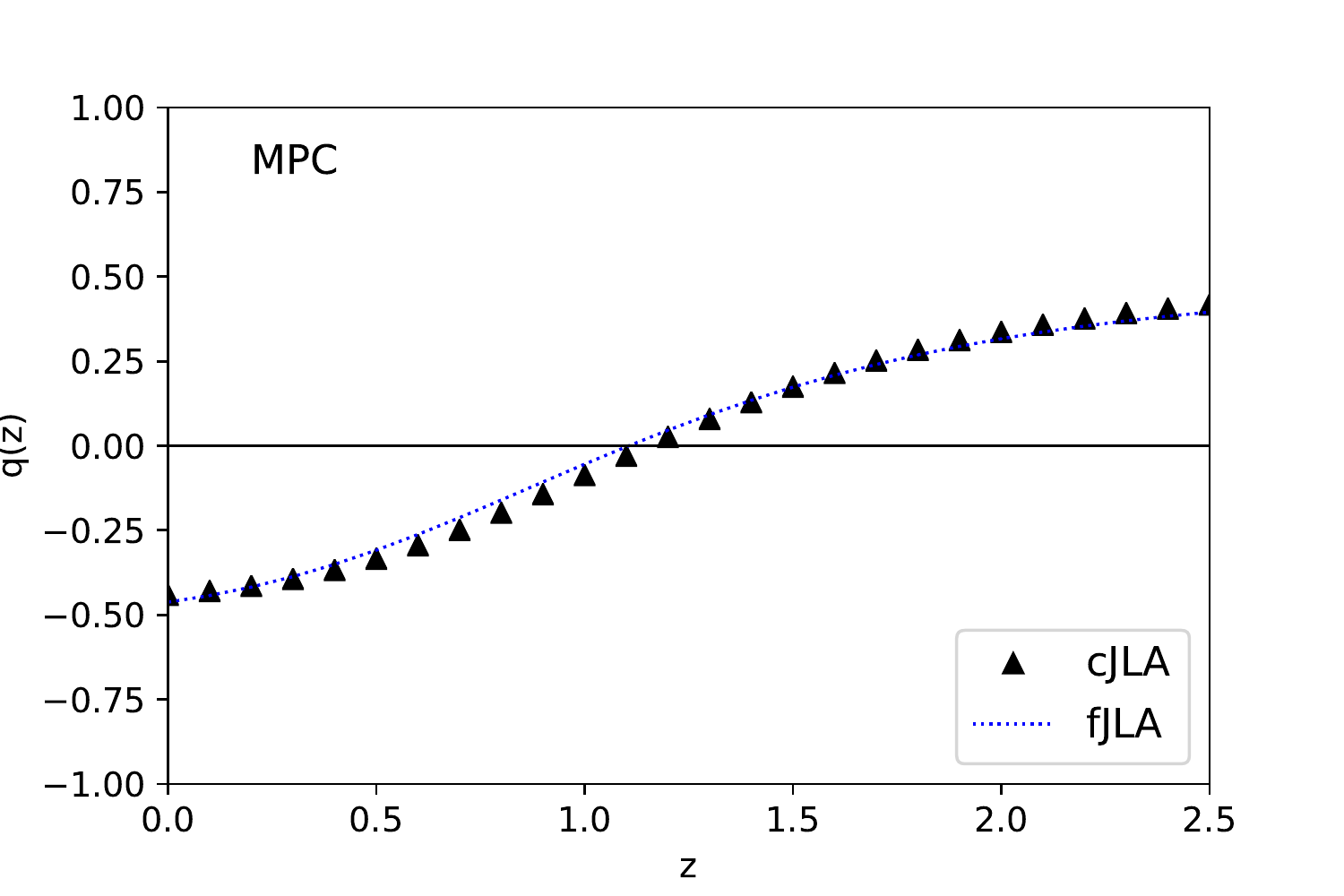} 
\caption{Reconstruction of the deceleration parameter $q(z)$ for the OC (top panel) and MPC (bottom panel) models using the constraints from the cJLA and fJLA samples when a flat prior on $h$ is considered. Notice that there is no significant differences in the $q(z)$ behavior using each SNIa sample.}
\label{fig:qzcvsf}
\end{figure}

The use of the fJLA sample to infer cosmological parameters has a high computational cost when several model are tested. To deal with this, we use the cJLA sample instead of the fJLA. Nevertheless, the former was computed under the standard cosmology. To asses how the Cardassian model constraints are biased when using each SNIa sample, we perform the parameter estimation with different combinations of  models, priors, and samples. The several constraints are presented in Tables \ref{tab:ocparnueva} and \ref{tab:MPCparnueva}. Notice that the mean values for the cosmological parameters in the OC model obtained from both SNIa samples are the same. For the MPC model, the largest difference is observed on the $l$ parameter (flat prior on $h$), $\sim0.18\sigma$. It is smaller for the $n$ parameter when employing a Gaussian prior on $h$.  Figure \ref{fig:fjlavscjla} illustrates the comparison of the confidence contours for these parameters using the cJLA and fJLA samples (flat prior on $h$). Figure \ref{fig:qzcvsf} shows that there is no significant difference in the reconstruction of the $q(z)$ parameter for the OC and MPC models using the constraints obtained from both SNIa samples. Therefore, to optimize the computational time, in the following analysis we only use the compressed JLA sample.

\subsection{The effects of the homogeneous OHD subsample in the parameter estimation.}
\begin{figure}
  \centering
          \includegraphics[width=6cm,scale=0.45]{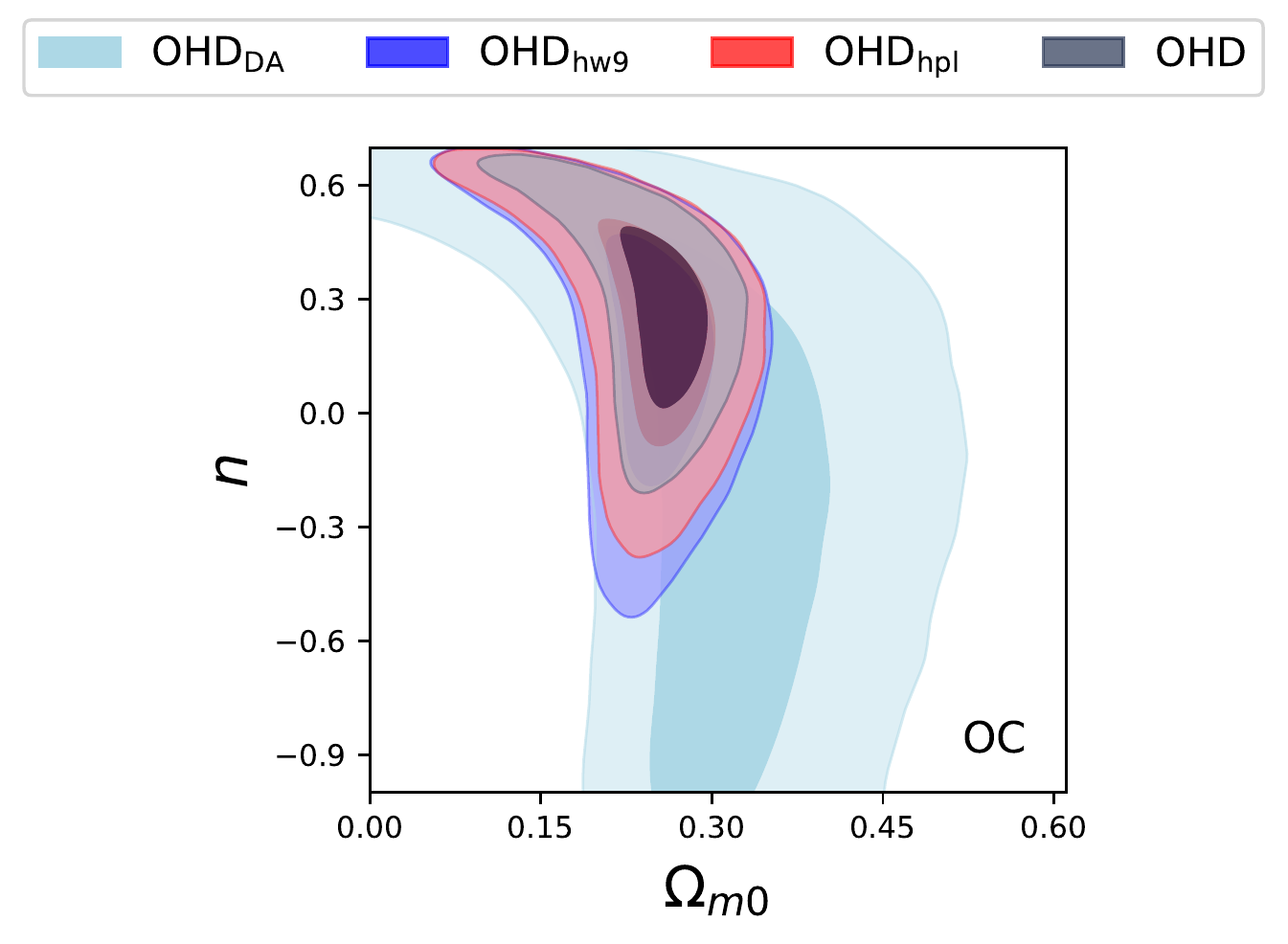} 
              \includegraphics[width=6cm,scale=0.45]{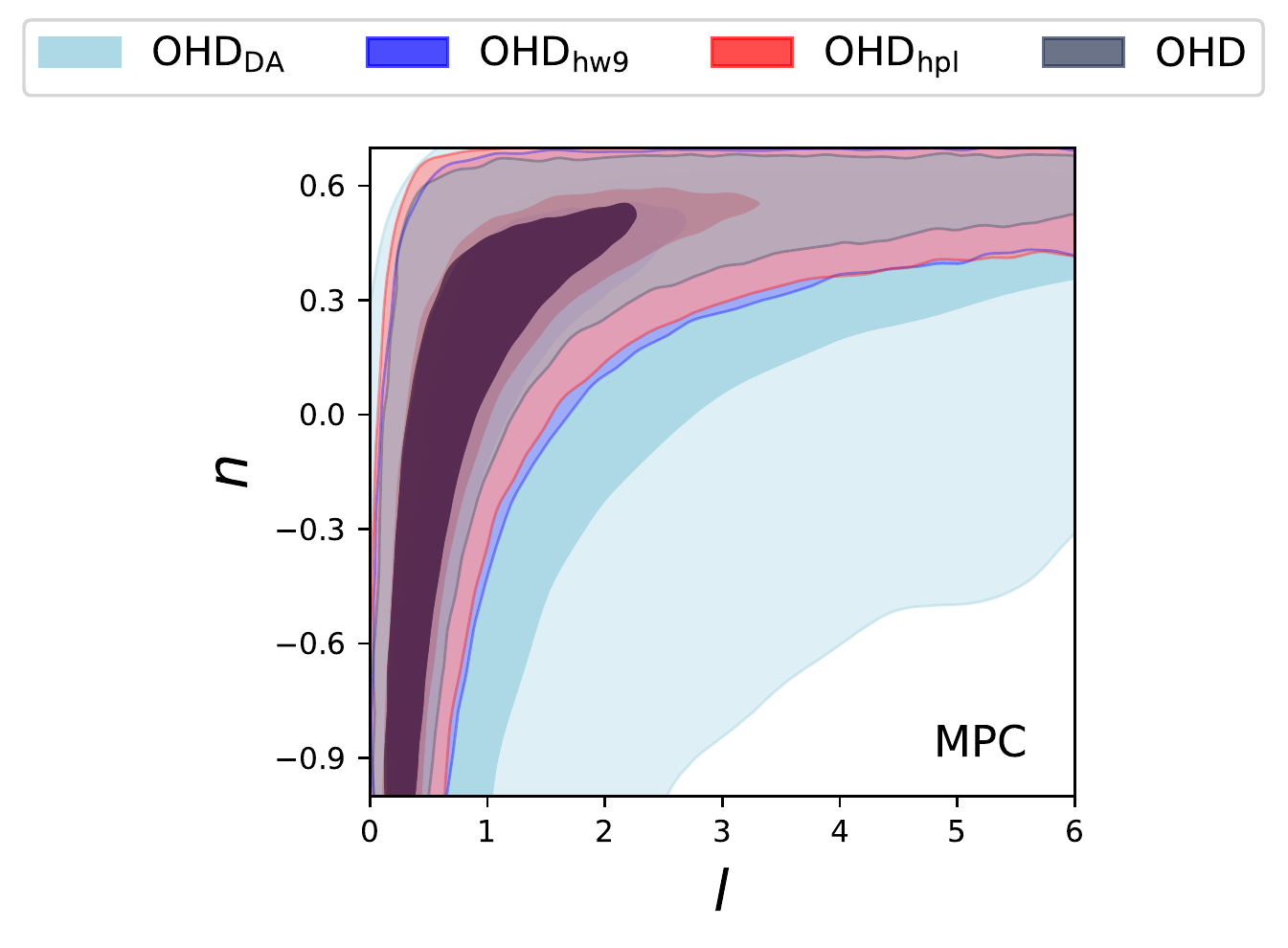} 
\caption{Confidence contours of the $\Omega_{m0}$-$n$ (top panel) and $n$-$l$ (bottom panel) constraints for the OC and MPC models within the $1\sigma$ and $3\sigma$ confidence levels using the OHD sample in Table \ref{tab:Hz}, the OHD$_{\mathrm{DA}}$ data set, and two samples containing the DA points plus those homogenized OHD points from clustering using the $r_{d}$ values from WMAP and Planck measurements. A flat prior on $h$ was considered in the parameter estimation.}
\label{fig:2DOC}
\end{figure}

\begin{figure}
  \centering
          \includegraphics[width=6cm,scale=0.45]{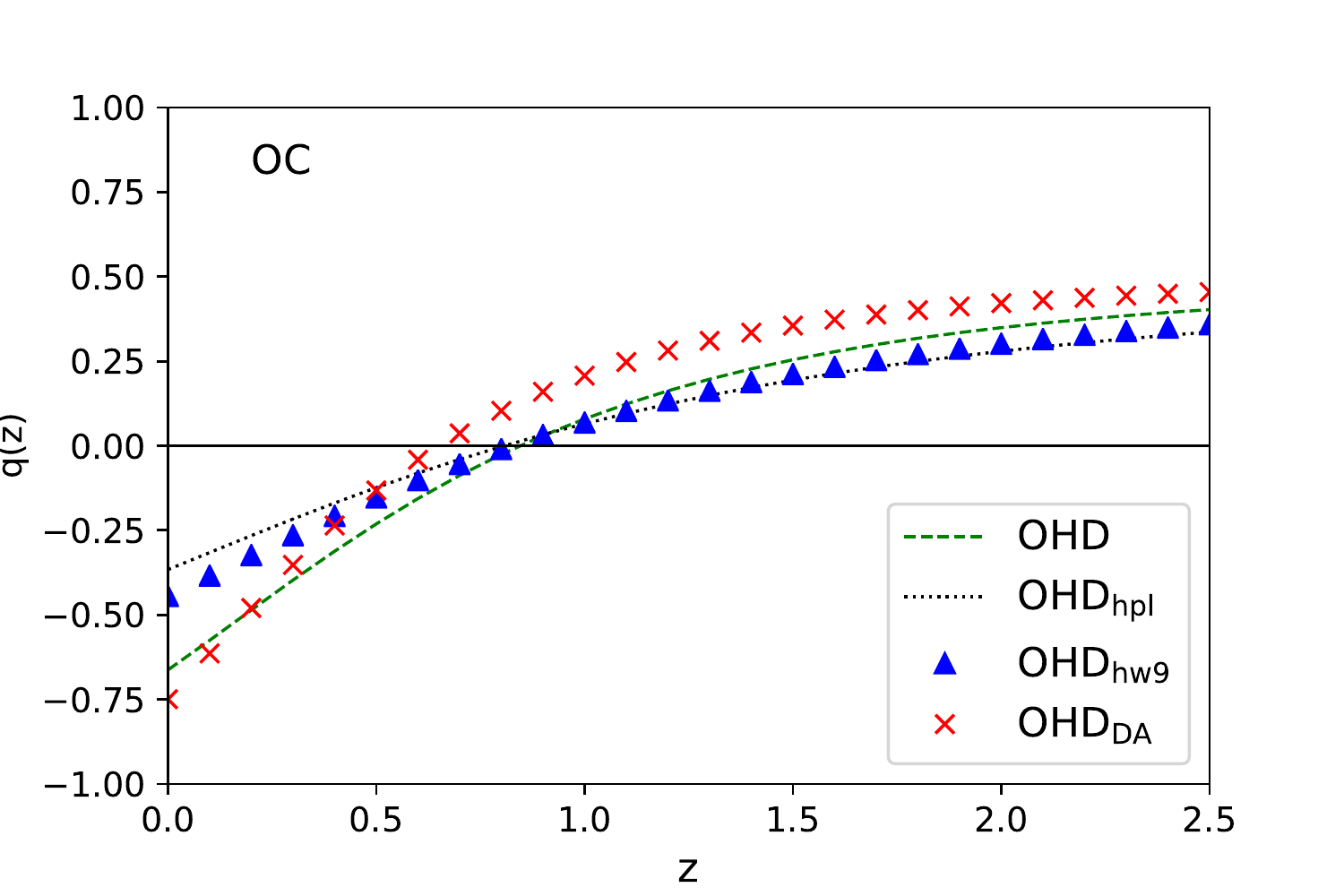} 
            \includegraphics[width=6cm,scale=0.45]{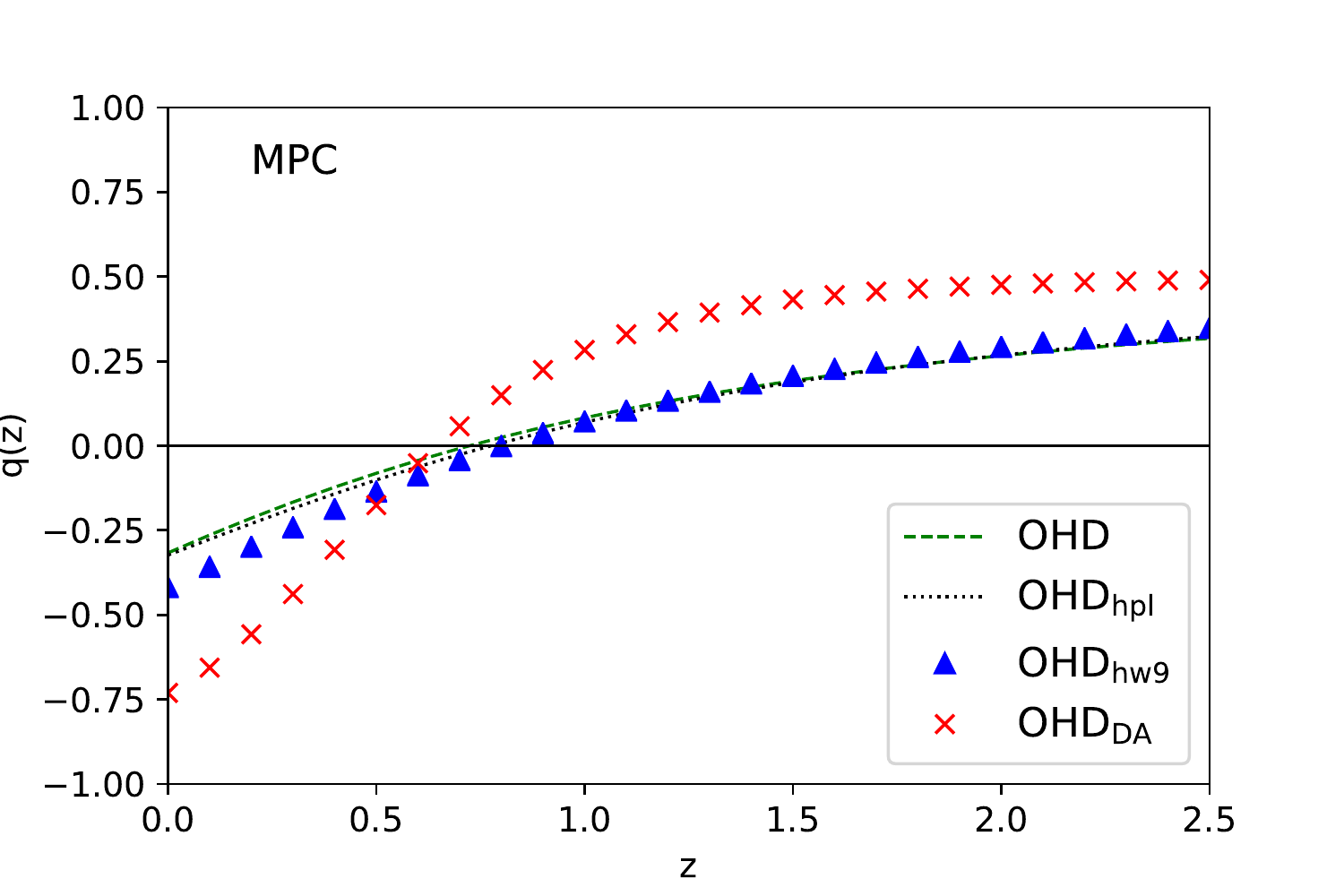} 
\caption{Reconstruction of the deceleration parameter $q(z)$ for the OC (top panel) and MPC (bottom panel) models using the constraints from different OHD samples when a flat prior on $h$ is considered.}
\label{fig:qzOHDs}
\end{figure}

In section \S \ref{subsubsec:homOHD}, an homogenized and model-independent OHD from clustering was constructed to avoid or reduce biased constraints due to the underlying cosmology or the underestimated systematic errors. Tables \ref{tab:ocparnueva}-\ref{tab:MPCparnueva} provide the OC and MPC bounds estimated from the combination of the new computed unbiased OHD from clustering with those obtained from the DA method. The increase on the error of $r_{d}$ also increases the error on $H(z)$, reducing the goodness of the fit ($\chi_{red}$). In spite of this, the advantage of these new limits is that they could be considered unbiased by different cosmological models. Figure \ref{fig:2DOC} shows the contours of the $\Omega_{m0}$-$n$ OC (top panel) and the $n$-$l$ MPC (bottom panel) parameters respectively using the different OHD samples. Note that all the bounds are consistent within the $1\sigma$ and $3\sigma$ C.L. Figure \ref{fig:qzOHDs} illustrates the $q(z)$ reconstructions using the different OHD data sets. Notice that for the OC model the homogenized OHD samples give slightly different $q(0)$ values than the obtained from the sample in Table \ref{tab:Hz}. For the MPC model, these differences are less significant.

\subsection{The effects of a different Gaussian prior on h.}
\begin{figure}
\centering
\includegraphics[width=5cm,scale=0.45]{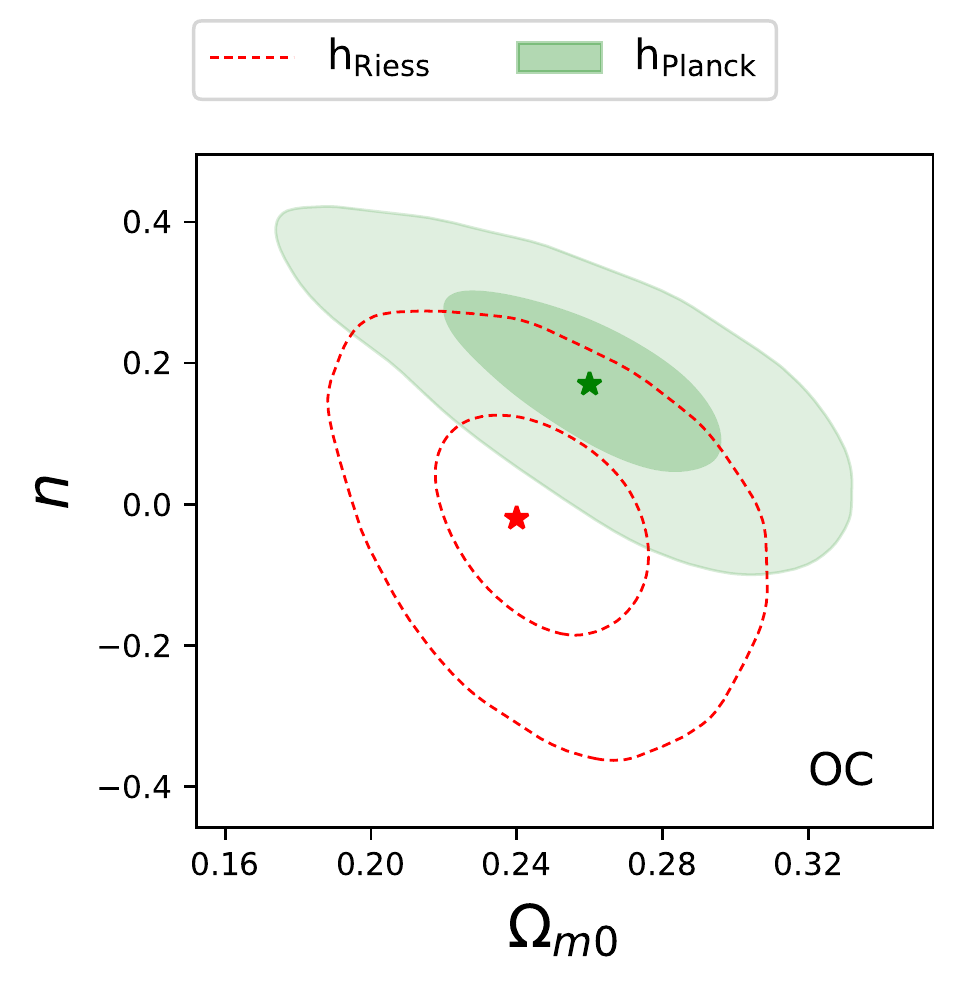} 
\includegraphics[width=5cm,scale=0.45]{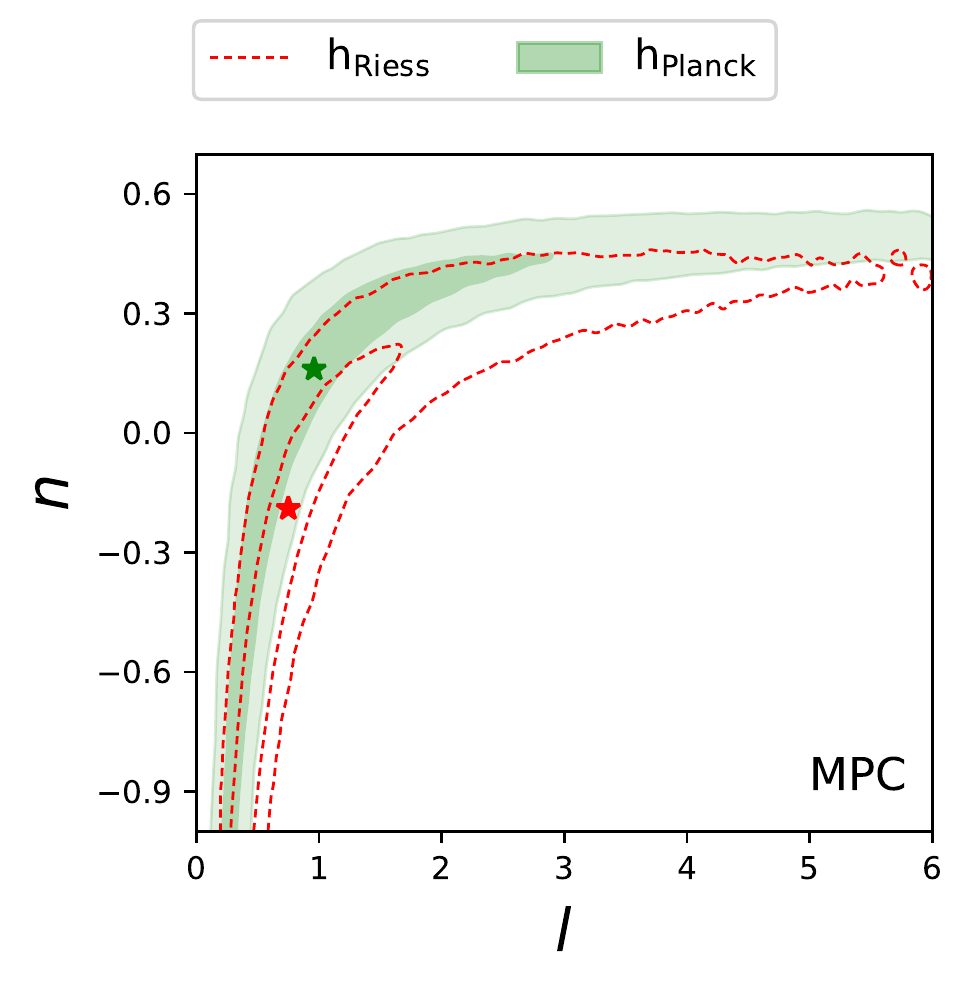} 
\caption{Comparison of the $\Omega_{m0}$-$n$ (top panel) and $l$-$n$ (bottom panel) confidence contours for the OC and MPC parameters respectively within the $1\sigma$ and $3\sigma$ confidence levels obtained from the OHD$_{\mathrm{hpl}}$ analysis using two Gaussian priors on $h$: $0.732\pm0.017$ \citep{Riess:2016jrr} and $0.678\pm0.009$ \citep{Planck2015XIII:2016}. The stars mark the mean values for each data set.}
\label{fig:CCpriors}
\end{figure}

 One of the most important problems in cosmology is the
 tension up to more than $3\sigma$ between the local measurements of the Hubble constant $H_{0}$ and those obtained from the CMB anisotropies \citep{Bernal:2016}. The latest estimation by the Planck collaboration \citep{Planck2015XIII:2016}, $h=0.678\pm 0.009$, is in disagreement with the first value given in Table \ref{tab:Hz}. Thus, using different Gaussian priors on $h$ will lead to different constraints on the OC and MPC parameters. Therefore, we carried out all our computations with both priors. Figure \ref{fig:CCpriors} illustrates how the confidence contours for the $\Omega_{m0}$-$n$ and $l$-$n$ parameters of the OC (top panel) and MPC (bottom panel) models obtained from OHD$_{\mathrm{hpl}}$ are shifted using each Gaussian prior. Although they are consistent at $3\sigma$, the tension in the constraints is important. In spite of these differences, both results drive the Universe to an accelerated phase but with slightly different transition redshifts (i.e. the redshift at which the Universe passes from a decelerated to an accelerated phase) and amplitude, $q(0)$. In addition, the OC and MPC bounds are consistent with the standard cosmology even when different Gaussian priors are considered. 

\begin{figure}
  \centering
          \includegraphics[width=6cm,scale=0.45]{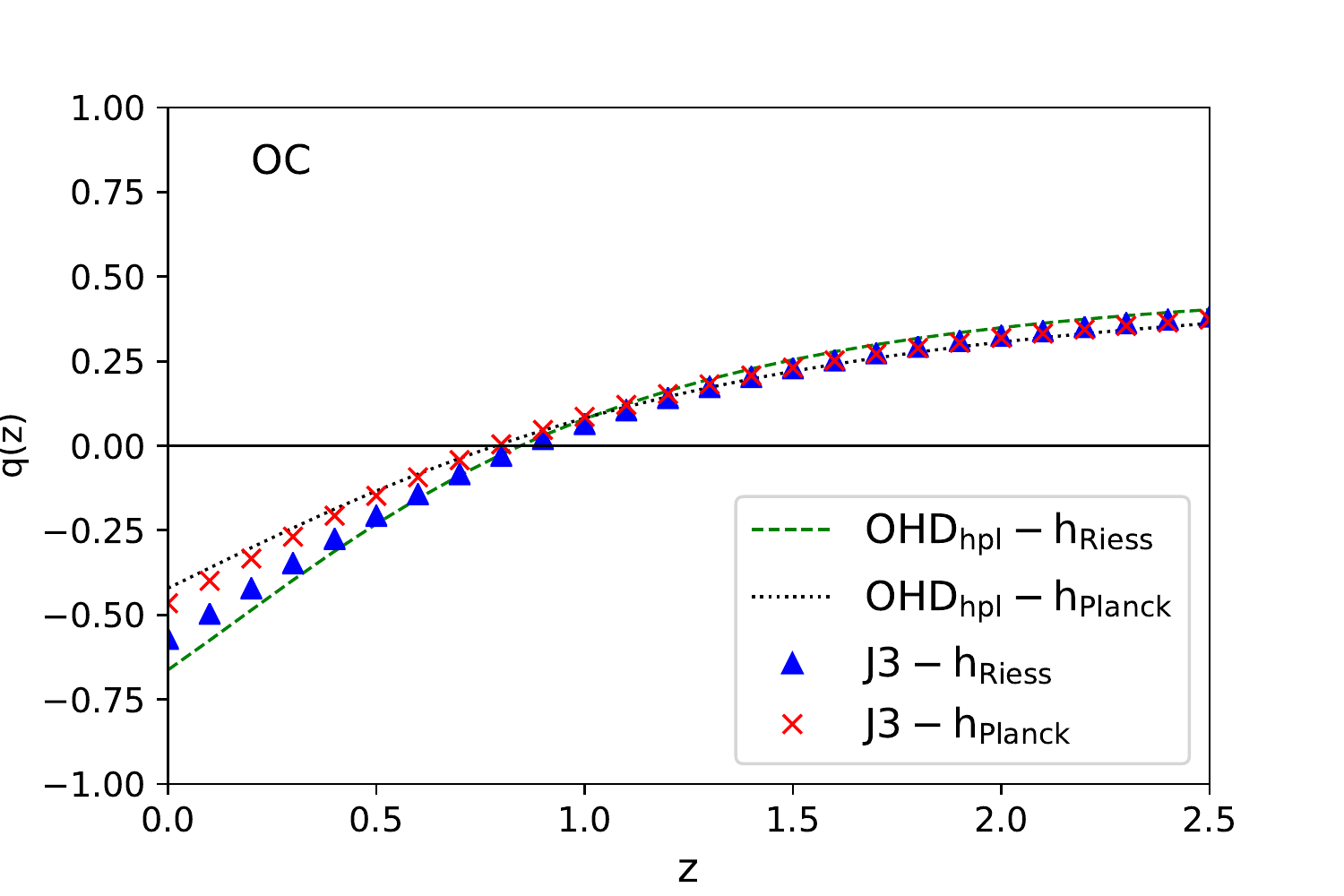} 
            \includegraphics[width=6cm,scale=0.45]{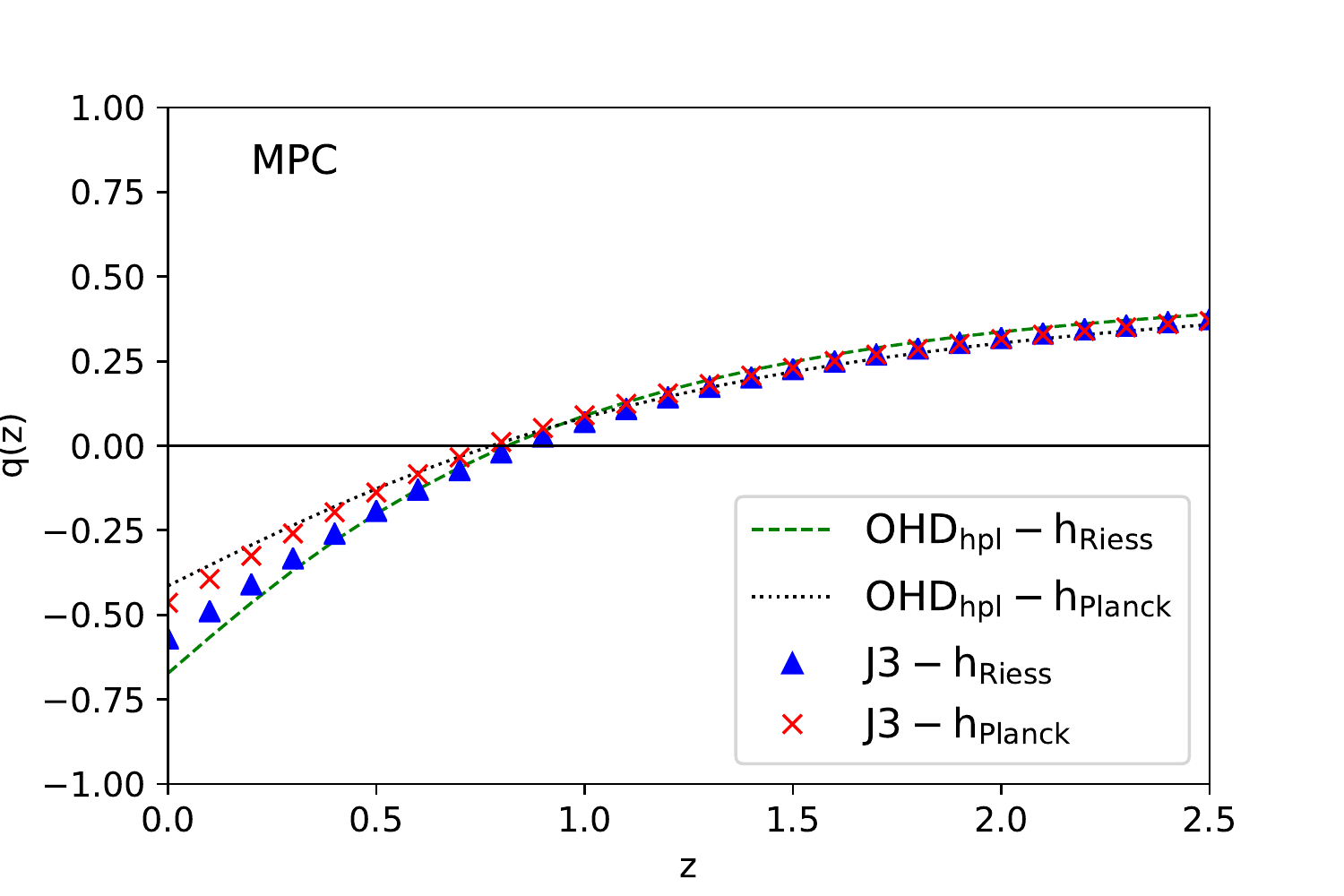} 
\caption{Reconstruction of the deceleration parameter $q(z)$ for the OC (top panel) and MPC (bottom panel) models using the constraints from the OHD$_{\mathrm{hpl}}$ sample and the joint analysis J3 when a different Gaussian prior on $h$ is considered: $0.732\pm0.017$ \citep{Riess:2016jrr} and $0.678\pm0.009$ \citep{Planck2015XIII:2016}} 
\label{fig:qzdifpriors}
\end{figure}

\subsection{Cosmological implications of the OC and MPC constraints}

\begin{figure*}
\centering
OC model\par\smallskip
\begin{tabular}{cc}
\subfloat[Flat prior on $h$]{\includegraphics[width=0.48\textwidth]{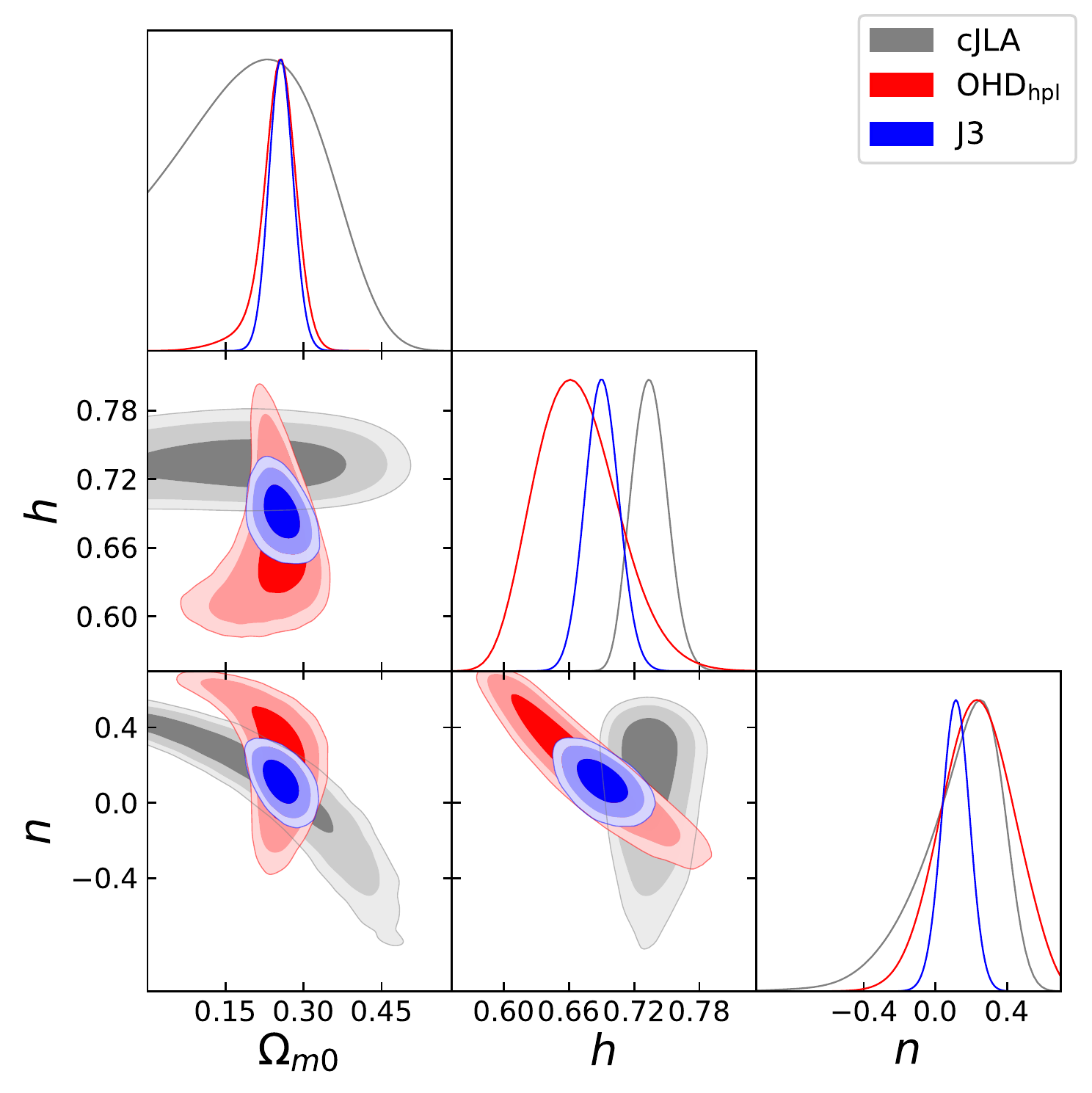}} & 
\subfloat[Gaussian prior on $h$]{\includegraphics[width=0.48\textwidth]{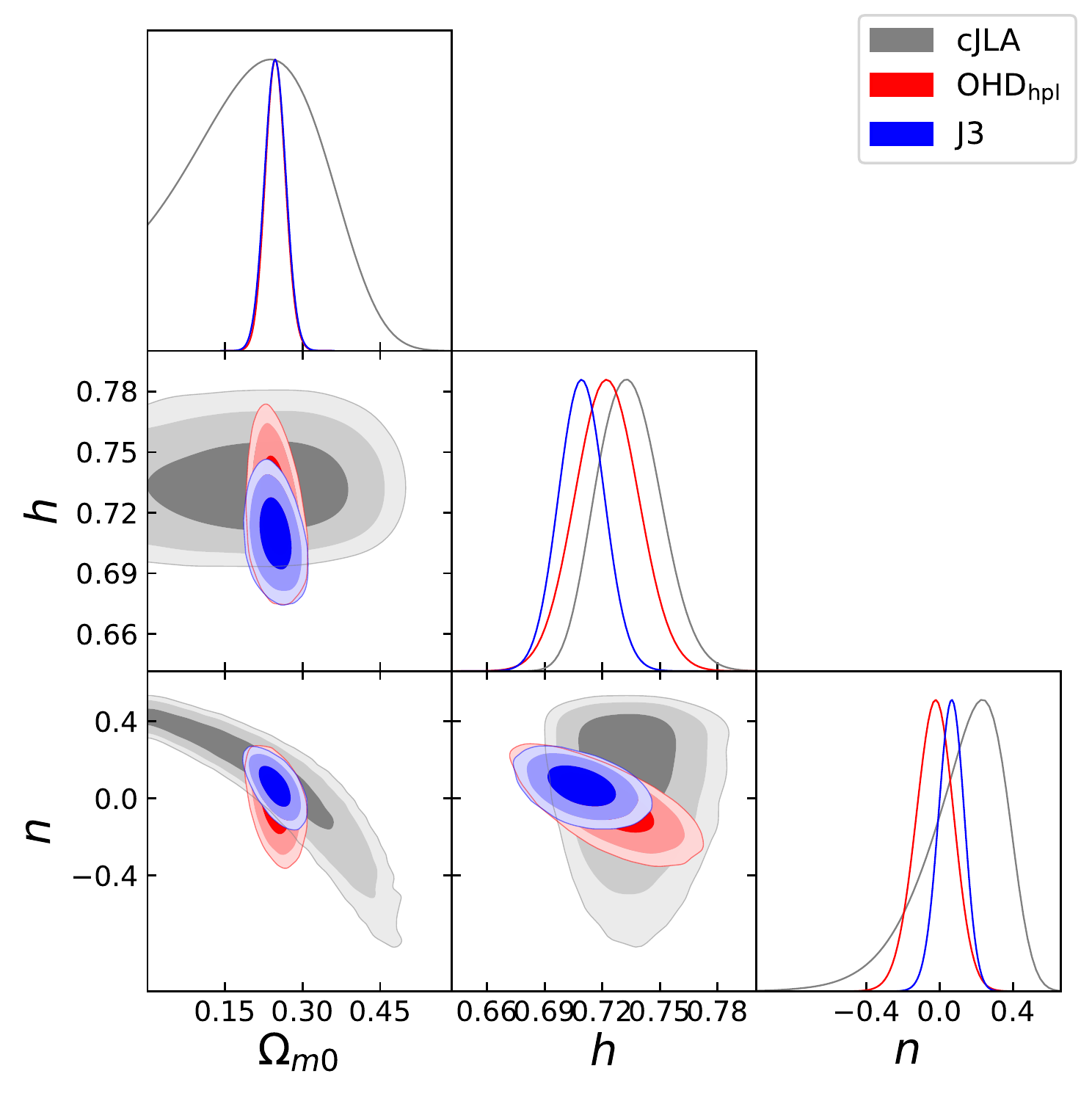}} \\
\end{tabular}
\caption{1D marginalized posterior distributions and the 2D $68\%$, $95\%$, $99.7\%$ confidence levels for the $\Omega_{m0}$, $h$, and $n$ parameters of the OC model assuming a flat and Gaussian (h$_\mathrm{Riess}$) prior on $h$}.
\label{fig:Cardcontours}
\end{figure*}

Figure \ref{fig:Cardcontours} shows the 1D marginalized posterior distributions and the 2D $68\%$, $95\%$, $99\%$ contours for the $\Omega_{m0}$, $h$, and $n$ parameters of the OC model obtained from OHD$_{\mathrm{hpl}}$, cJLA, and J3 with flat (left panel) and Gaussian (right panel) priors on $h$.
Assuming a flat prior on $h$, the $\Omega_{m0}$, $h$ constraints obtained from the different data sets are consistent between them and are in agreement with Planck measurements for the standard model. For the $n$ parameter we found a tension in the constraints obtained from the different data sets. Nevertheless, the bounds have large uncertainties and are consistent among them within the $1\sigma$ CL. 
Our $n$ constraints are consistent within the $1\sigma$ CL with those
estimated by other authors, for instance, $n=-0.04^{+0.07}_{-0.07}$ \citep{Xu:2012ig}, $n=0.16^{+0.30}_{-0.52}$ \citep{Wei:2015kua}, and $n=-0.022^{+0.05}_{-0.05}$ \citep{Zhai:2017}.
It is worth to note that, when the cJLA data are used, $\Omega_{m}$ drop at extremely low values (see the $\Omega_{m0}$-$n$ contour), which is consistent
with the results by \citet{Wei:2015kua} who obtained a similar contour using the Union 2.1 data set. In addition, the $\chi_{red}^{2}$ values 
from the SN Ia data suggest that their errors (cJLA sample) are underestimated.

\begin{figure}
\centering
\includegraphics[width=7cm,scale=0.45]{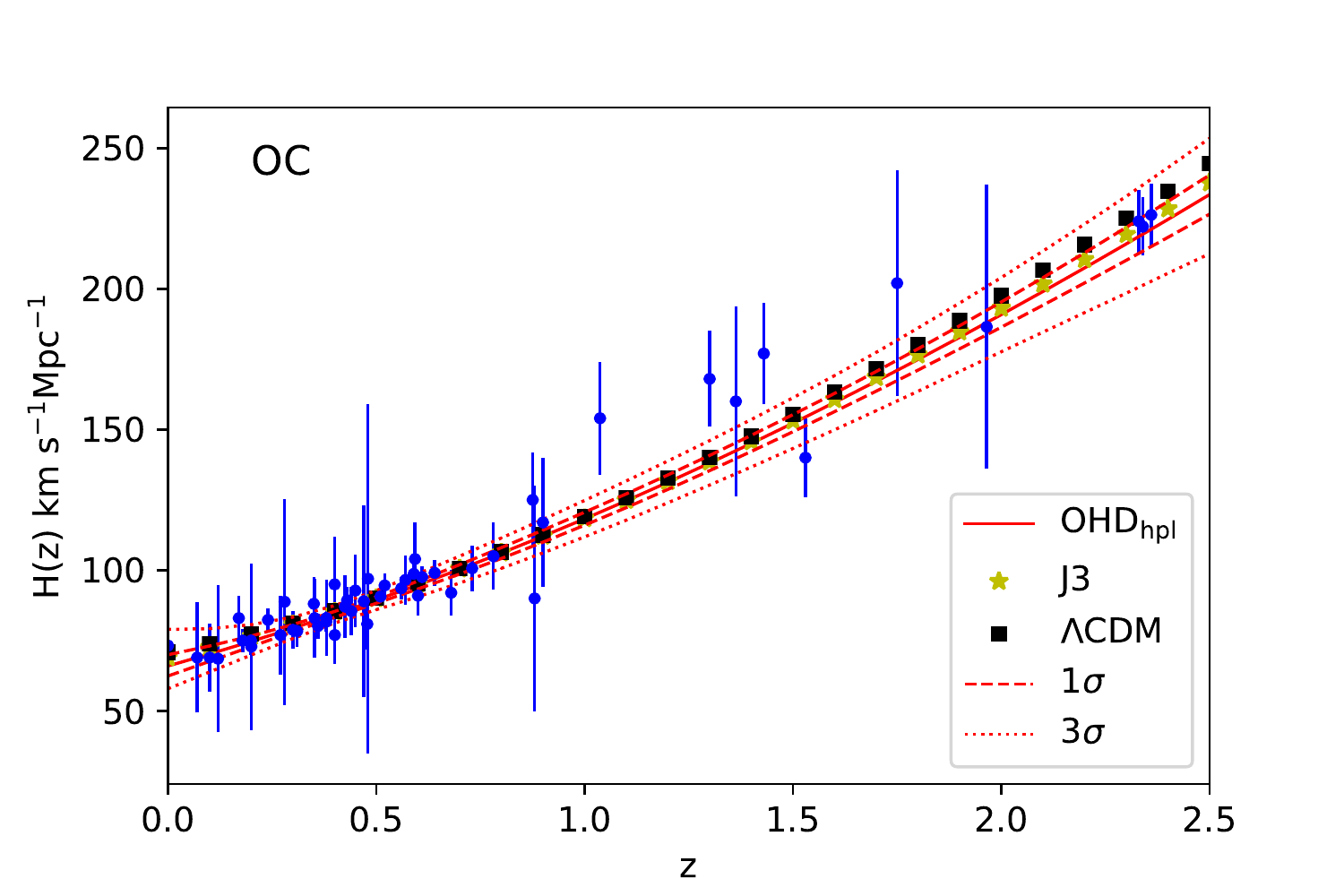} 
\includegraphics[width=7cm,scale=0.45]{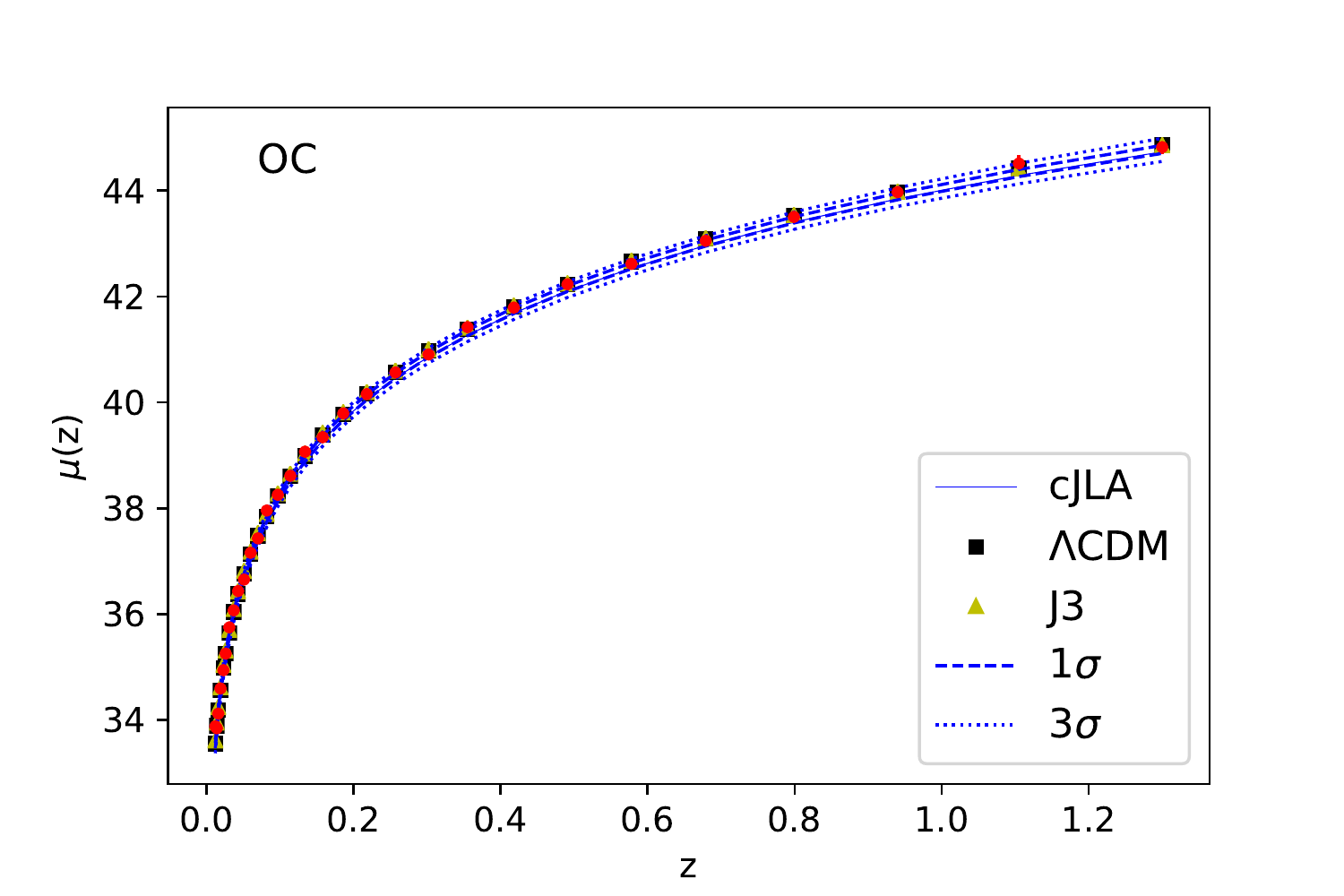} 
\caption{Fitting to OHD$_{\mathrm{hpl}}$ (top panel) and cJLA data (bottom panel) using the mean values from the OHD$_{\mathrm{hpl}}$ (red solid lines), cJLA (blue solid lines) and J3 (yellow star and triangle) analysis for $\Lambda$CDM model (black squares) and OC model with a flat prior on $h$. The dashed and the dotted lines represent the $68\%$ and $99.7\%$ confidence levels respectively.}
\label{fig:hzsnoc}
\end{figure}

On the other hand, when the Gaussian prior on $h$ by \citet{Riess:2016jrr} is considered, the OHD$_{\mathrm{hpl}}$  provides a better fitting for the OC parameters than those obtained when a flat prior is used (see the $\chi_{red}^{2}$ values). SN Ia data show no important statistical difference in the parameter estimation when flat or Gaussian priors are employed. 
Notice that we obtain stringent constraints from the joint analysis (see Fig. \ref{fig:Cardcontours}), which prefers values around $n\sim 0$. Figure \ref{fig:hzsnoc} shows the fittings to the OHD$_{\mathrm{hpl}}$ (top panel) and cJLA data (bottom panel) using the OHD$_{\mathrm{hpl}}$, cJLA and J3 constraints for the OC model.  A Monte Carlo approach was performed to propagate the error on the $1\sigma$, and $3\sigma$ CL.
The comparison between these results and the $\Lambda$CDM fitting reveals that both models are in agreement with the data and there is no significant difference between them. In addition, when the J1, J2, and J4 constraints are used, we found consistent results within the $1\sigma$ confidence level. Therefore, the extra term in the Eq. (\ref{C1}) to the canonical Friedmann equation acts like a CC. However, in the OC models this term can be sourced by an extra dimension instead of the expected vacuum energy. 

\begin{table}
\caption{Mean values for the OC parameters ($\Omega_{m0}$, $h$, $n$) derived from OHD and SN Ia  data of the cJLA and fJLA sample.}
\centering
\resizebox{0.48\textwidth}{!}{
\begin{tabular}{|ccccccc|}
\multicolumn{7}{c}{OC model}\\
\hline
Parameter & OHD & OHD$_{\mathrm{DA}}$ & OHD$_{\mathrm{hpl}}$ & OHD$_{\mathrm{hw9}}$ & cJLA& fJLA\\
\hline
\multicolumn{7}{|c|}{Flat prior on $h$}\\
$\chi^{2}_{min}$ & 25.37 & 15.22  & 21.25  & 22.52  & 32.95 &682.28\\
 $\chi^{2}_{red}$ & 0.52 & 0.54 & 0.44 & 0.46 & 1.22 &0.93\\
 $\Omega_{m0}$ & $0.25^{+0.02}_{-0.02}$ & $0.30^{+0.06}_{-0.06}$ & $0.25^{+0.02}_{-0.03}$ & $0.25^{+0.03}_{-0.03}$  & $0.22^{+0.11}_{-0.12}$  &$0.22^{+0.11}_{-0.12}$\\
$h$ & $0.65^{+0.03}_{-0.03}$ & $0.69^{+0.06}_{-0.05}$ & $0.66^{+0.04}_{-0.03}$ & $0.66^{+0.04}_{-0.03}$  & $0.72^{+0.19}_{-0.19}$ &$0.72^{+0.18}_{-0.18}$\\
$n$ & $0.26^{+0.16}_{-0.15}$ & $-0.19^{+0.51}_{-0.50}$  & $0.23^{+0.20}_{-0.20}$  & $0.16^{+0.22}_{-0.22}$ & $0.16^{+0.17}_{-0.26}$ & $0.16^{+0.18}_{-0.26}$ \\
$M(M_{B}^{1})$ & -- & -- & -- & -- & $0.07^{+0.5}_{-0.66}$ & $-18.96^{+0.49}_{-0.64}$ \\
$\Delta_{M}$ & -- & -- & -- & -- & -- & $-0.06^{+0.02}_{-0.01}$ \\
$\alpha$ & -- & -- & -- & -- & -- & $0.14^{+0.006}_{-0.006}$ \\
$\beta$ & -- & -- & -- & -- & -- & $3.10^{+0.08}_{-0.07}$\\
\multicolumn{7}{|c|}{Gaussian prior on $h=0.732\pm 0.017$}\\
$\chi^{2}_{min}$ & 28.86 & 14.47  & 22.83 & 23.91 & 32.95 &682.28\\
$\chi^{2}_{red}$ & 0.60  & 0.51  & 0.47 & 0.49 & 1.22 &0.93\\
$\Omega_{m0}$ & $0.24^{+0.01}_{-0.01}$ & $0.31^{+0.03}_{-0.04}$ & $0.24^{+0.01}_{-0.01}$ & $0.24^{+0.01}_{-0.01}$ & $0.22^{+0.11}_{-0.12}$ &$0.22^{+0.11}_{-0.12}$\\
$h$ & $0.71^{+0.01}_{-0.01}$ & $0.72^{+0.01}_{-0.01}$  & $0.72^{+0.01}_{-0.01}$  & $0.72^{+0.01}_{-0.01}$  & $0.73^{+0.01}_{-0.01}$ &$0.73^{+0.01}_{-0.01}$\\
$n$ & $-0.01^{+0.08}_{-0.08}$ & $-0.43^{+0.28}_{-0.30}$  & $-0.02^{+0.09}_{-0.10}$  & $-0.11^{+0.10}_{-0.11}$  & $0.16^{+0.17}_{-0.26}$ & $0.16^{+0.18}_{-0.26}$ \\
$M(M_{B}^{1})$ & -- & -- & -- & -- & $0.10^{+0.05}_{-0.05}$ & $-18.94^{+0.05}_{-0.05}$ \\
$\Delta_{M}$ &-- & -- & -- & -- &-- & $-0.06^{+0.02}_{-0.01}$ \\
$\alpha$ &-- & -- &  --& --  &-- & $0.14^{+0.006}_{-0.006}$ \\
$\beta$ &-- & -- & -- & -- & --& $3.10^{+0.08}_{-0.07}$ \\
\multicolumn{7}{|c|}{Gaussian prior on $h=0.678\pm 0.009$}\\
$\chi^{2}_{min}$ & 25.24 & 14.53  & 20.79 & 22.04 & 32.95 &--\\
$\chi^{2}_{red}$ & 0.52  & 0.51  & 0.43 & 0.45 & 1.22 &--\\
$\Omega_{m0}$ & $0.26^{+0.01}_{-0.01}$ & $0.33^{+0.05}_{-0.07}$ & $0.26^{+0.02}_{-0.02}$ & $0.25^{+0.02}_{-0.02}$ & $0.22^{+0.11}_{-0.12}$ &--\\
$h$ & $0.67^{+0.008}_{-0.008}$ & $0.67^{+0.009}_{-0.009}$  & $0.67^{+0.008}_{-0.008}$  & $0.67^{+0.008}_{-0.008}$  & $0.67^{+0.009}_{-0.009}$ &--\\
$n$ & $0.15^{+0.06}_{-0.06}$ & $-0.05^{+0.27}_{-0.32}$  & $0.17^{+0.08}_{-0.08}$  & $0.09^{+0.08}_{-0.09}$  & $0.16^{+0.17}_{-0.26}$ & -- \\
$M$ & -- & -- & -- & -- & $-0.06^{+0.03}_{-0.03}$ & -- \\
\hline
\end{tabular}}
\label{tab:ocparnueva}
\end{table}

\begin{table}
\caption{Mean values for the MPC parameters ($\Omega_{m0}$, $h$, $n$, $l$) derived from OHD and SN Ia  data of the cJLA and fJLA sample.}
\centering
\resizebox{0.48\textwidth}{!}{
\begin{tabular}{|ccccccc|}
\multicolumn{7}{c}{MPC model}\\
\hline
Parameter & OHD & OHD$_{\mathrm{DA}}$ & OHD$_{\mathrm{hpl}}$ & OHD$_{\mathrm{hw9}}$ &cJLA& fJLA\\
\hline
\multicolumn{7}{|c|}{Flat prior on $h$}\\
$\chi^{2}_{min}$ & 25.31 & 17.95  & 21.17  & 22.98  & 33.76 &682.92\\
$\chi^{2}_{red}$ & 0.53 & 0.66 & 0.45 & 0.48 & 1.29 &0.93\\
$\Omega_{m0}$ & $0.25^{+0.04}_{-0.04}$ & $0.32^{+0.06}_{-0.07}$ & $0.25^{+0.04}_{-0.05}$ & $0.25^{+0.04}_{-0.04}$  & $0.22^{+0.12}_{-0.13}$  &$0.22^{+0.12}_{-0.13}$\\
$h$ & $0.64^{+0.03}_{-0.02}$ & $0.68^{+0.07}_{-0.05}$ & $0.65^{+0.03}_{-0.03}$ & $0.65^{+0.04}_{-0.03}$  & $0.72^{+0.18}_{-0.19}$ &$0.72^{+0.18}_{-0.18}$\\
$n$ & $0.17^{+0.34}_{-0.68}$ & $0.10^{+0.38}_{-0.60}$  & $0.25^{+0.29}_{-0.68}$  & $0.15^{+0.34}_{-0.65}$ & $0.36^{+0.07}_{-0.33}$ & $0.33^{+0.09}_{-0.50}$ \\
$l$ & $0.77^{+1.45}_{-0.43}$ & $2.13^{+2.34}_{-1.33}$  & $0.95^{+1.90}_{-0.58}$  & $0.92^{+1.66}_{-0.52}$ & $2.61^{+2.27}_{-1.83}$ & $2.09^{+2.51}_{-1.49}$  \\
$M(M_{B}^{1})$ & -- & -- & -- & -- & $0.08^{+0.50}_{-0.67}$ & $-18.97^{+0.49}_{-0.65}$ \\
$\Delta_{M}$ & -- & -- & -- & -- & -- & $-0.06^{+0.02}_{-0.01}$ \\
$\alpha$ & -- & -- & -- & -- & -- & $0.14^{+0.006}_{-0.006}$ \\
$\beta$ & -- & -- & -- & -- & -- & $3.10^{+0.08}_{-0.07}$\\
\multicolumn{7}{|c|}{Gaussian prior on $h=0.732\pm 0.017$}\\
$\chi^{2}_{min}$ & 27.75 & 14.92  & 22.40 & 23.42 & 33.75 & 683.17 \\
$\chi^{2}_{red}$ & 0.59  & 0.55  & 0.47 & 0.49 & 1.29 &0.93\\
$\Omega_{m0}$ & $0.24^{+0.01}_{-0.01}$ & $0.32^{+0.03}_{-0.04}$ & $0.24^{+0.02}_{-0.02}$ & $0.23^{+0.02}_{-0.02}$ & $0.22^{+0.12}_{-0.14}$ &$0.22^{+0.12}_{-0.13}$\\
$h$ & $0.71^{+0.01}_{-0.01}$ & $0.72^{+0.01}_{-0.01}$  & $0.72^{+0.01}_{-0.01}$  & $0.72^{+0.01}_{-0.01}$  & $0.73^{+0.01}_{-0.01}$ &$0.73^{+0.01}_{-0.01}$\\
$n$ & $-0.34^{+0.40}_{-0.42}$ & $-0.03^{+0.24}_{-0.49}$  & $-0.19^{+0.39}_{-0.50}$  & $-0.28^{+0.39}_{-0.46}$  & $0.36^{+0.07}_{-0.33}$ & $0.34^{+0.08}_{-0.47}$ \\
$l$ & $0.62^{+0.49}_{-0.20}$ & $2.12^{+2.29}_{-1.20}$  & $0.75^{+0.80}_{-0.30}$  & $0.77^{+0.73}_{-0.28}$ & $2.60^{+2.27}_{-1.81}$ & $2.26^{+2.44}_{-1.63}$ \\
$M(M_{B}^{1})$ & -- & -- & -- & --  &  $0.10^{+0.05}_{-0.05}$& $-18.94^{+0.05}_{-0.05}$ \\
$\Delta_{M}$ & --& -- & -- & -- & -- & $-0.06^{+0.02}_{-0.01}$ \\
$\alpha$ & -- & -- & --  &  -- & -- & $0.14^{+0.006}_{-0.006}$ \\
$\beta$ & -- &  -- & -- & -- & -- & $3.10^{+0.08}_{-0.07}$ \\
\multicolumn{7}{|c|}{Gaussian prior on $h=0.678\pm 0.009$}\\
$\chi^{2}_{min}$ & 25.03 & 14.70  & 20.84 & 21.96 & 33.79 & -- \\
$\chi^{2}_{red}$ & 0.53  & 0.54  & 0.44 & 0.46 & 1.29 &--\\
$\Omega_{m0}$ & $0.25^{+0.02}_{-0.02}$ & $0.35^{+0.05}_{-0.08}$ & $0.26^{+0.03}_{-0.03}$ & $0.25^{+0.02}_{-0.03}$ & $0.22^{+0.12}_{-0.13}$ & --\\
$h$ & $0.67^{+0.008}_{-0.008}$ & $0.67^{+0.009}_{-0.009}$  & $0.67^{+0.009}_{-0.009}$  & $0.67^{+0.009}_{-0.008}$  &$0.67^{+0.009}_{-0.009}$  & --\\
$n$ & $-0.01^{+0.35}_{-0.57}$ & $0.24^{+0.19}_{-0.53}$  & $0.16^{+0.26}_{-0.60}$  & $0.03^{+0.31}_{-0.58}$  & $0.36^{+0.07}_{-0.36}$  & -- \\
$l$ & $0.71^{+0.95}_{-0.34}$ & $2.08^{+2.44}_{-1.37}$  & $0.96^{+1.55}_{-0.55}$  & $0.87^{+1.22}_{-0.45}$ & $2.59^{+2.27}_{-1.8}$  & -- \\
$M$ & -- & -- & -- & -- & $-0.06^{+0.03}_{-0.03}$  & -- \\
\hline
\end{tabular}}
\label{tab:MPCparnueva}
\end{table}

\begin{table}
\caption{Mean values for the OC parameters ($\Omega_{m0}$, $h$, $n$) derived from a joint analysis OHD$+$cJLA.}
\centering
\resizebox{0.48\textwidth}{!}{
\begin{tabular}{|ccccccc|}
\multicolumn{7}{c}{OC model}\\
\hline
Data set & $\chi^{2}_{min}$ & $\chi^{2}_{red}$& $\Omega_{m0}$ & $h$ &$n$&M\\
\hline
\multicolumn{7}{|c|}{Flat prior on $h$}\\
J1 & 58.91 & $0.71$ & $0.25^{+0.01}_{-0.01}$ & $0.68^{+0.01}_{-0.01}$ & $0.12^{+0.06}_{-0.06}$&$-0.04^{+0.03}_{-0.03}$\\
J2 & 48.28 & $0.58$ & $0.30^{+0.05}_{-0.05}$ & $0.68^{+0.02}_{-0.02}$ & $-0.001^{+0.15}_{-0.17}$& $-0.03^{+0.06}_{-0.06}$\\
J3 & 54.28 & $0.66$ & $0.25^{+0.02}_{-0.02}$ & $0.69^{+0.01}_{-0.01}$ & $0.11^{+0.07}_{-0.07}$&$-0.02^{+0.04}_{-0.04}$\\
J4 & 55.17 & $0.67$ & $0.25^{+0.02}_{-0.02}$ & $0.67^{+0.01}_{-0.01}$ & $0.10^{+0.07}_{-0.08}$&$-0.07^{+0.04}_{-0.04}$\\
\multicolumn{7}{|c|}{Gaussian prior on $h=0.732\pm0.017$}\\
J1 & 63.34 & $0.76$ & $0.25^{+0.01}_{-0.01}$ & $0.70^{+0.01}_{-0.01}$ & $0.05^{+0.05}_{-0.05}$&$0.001^{+0.02}_{-0.03}$\\
J2 & 50.73 & $0.61$ & $0.27^{+0.04}_{-0.05}$ & $0.71^{+0.01}_{-0.01}$ & $0.001^{+0.14}_{-0.15}$ &$0.03^{+0.04}_{-0.04}$\\
J3 & 57.37 & $0.69$ & $0.24^{+0.02}_{-0.02}$ & $0.70^{+0.01}_{-0.01}$ & $0.06^{+0.06}_{-0.07}$&$0.01^{+0.03}_{-0.03}$\\
J4 & 60.96 & $0.73$ & $0.24^{+0.02}_{-0.02}$ & $0.70^{+0.01}_{-0.01}$ & $0.04^{+0.07}_{-0.07}$&$-0.01^{+0.03}_{-0.03}$\\
\multicolumn{7}{|c|}{Gaussian prior on $h=0.678\pm0.009$}\\
J1 & 59.04 & $0.71$ & $0.26^{+0.01}_{-0.01}$ & $0.67^{+0.007}_{-0.007}$ & $0.13^{+0.05}_{-0.05}$&$-0.05^{+0.02}_{-0.02}$\\
J2 & 48.53 & $0.58$ & $0.31^{+0.04}_{-0.05}$ & $0.67^{+0.008}_{-0.008}$ & $0.001^{+0.15}_{-0.17}$ &$0.05^{+0.03}_{-0.03}$\\
J3 & 54.80 & $0.66$ & $0.26^{+0.02}_{-0.02}$ & $0.68^{+0.007}_{-0.007}$ & $0.13^{+0.06}_{-0.07}$&$-0.04^{+0.02}_{-0.02}$\\
J4 & 55.18 & $0.65$ & $0.25^{+0.02}_{-0.02}$ & $0.67^{+0.007}_{-0.007}$ & $0.10^{+0.07}_{-0.07}$&$-0.06^{+0.02}_{-0.02}$\\
\hline
\end{tabular}}
\label{tab:ocparjoint}
\end{table}

\begin{table}
\caption{Mean values for the MPC parameters ($\Omega_{m0}$, $h$, $n$, $l$) derived from a joint analysis OHD$+$cJLA. }
\centering
\resizebox{0.48\textwidth}{!}{
\begin{tabular}{|cccccccc|}
\multicolumn{8}{c}{MPC model}\\
\hline
Data set & $\chi^{2}_{min}$ & $\chi^{2}_{red}$& $\Omega_{m0}$ & $h$ &$n$ &$l$& M\\
\hline
\multicolumn{8}{|c|}{Flat prior on $h$}\\
J1 & 58.61 & $0.71$ & $0.25^{+0.02}_{-0.02}$ & $0.68^{+0.01}_{-0.01}$ & $-0.03^{+0.34}_{-0.56}$& $0.74^{+0.90}_{-0.35}$ &$-0.04^{+0.03}_{-0.03}$\\
J2 & 48.25 & $0.58$ & $0.32^{+0.05}_{-0.07}$ & $0.68^{+0.02}_{-0.02}$ & $0.25^{+0.13}_{-0.51}$& $2.00^{+2.244}_{-1.33}$& $-0.03^{+0.06}_{-0.06}$\\
J3 & 54.23 & $0.66$ & $0.25^{+0.03}_{-0.03}$ & $0.68^{+0.01}_{-0.01}$ & $0.06^{+0.29}_{-0.58}$& $0.89^{+1.29}_{-0.47}$&$-0.02^{+0.04}_{-0.04}$\\
J4& 55.14 & $0.67$ & $0.25^{+0.03}_{-0.03}$ & $0.67^{+0.01}_{-0.01}$ & $0.06^{+0.28}_{-0.58}$& $0.91^{+1.29}_{-0.49}$&$-0.07^{+0.04}_{-0.04}$\\
\multicolumn{8}{|c|}{Gaussian prior on $h=0.732\pm0.017$}\\
J1 & 62.91 & $0.75$ & $0.24^{+0.02}_{-0.02}$ & $0.70^{+0.01}_{-0.01}$ & $-0.14^{+0.36}_{-0.51}$& $0.70^{+0.72}_{-0.29}$&$-0.0006^{+0.03}_{-0.03}$\\
J2 & 50.81 & $0.61$ & $0.30^{+0.04}_{-0.06}$ & $0.71^{+0.01}_{-0.01}$ & $0.22^{+0.14}_{-0.54}$& $1.77^{+2.44}_{-1.17}$&$0.03^{+0.04}_{-0.04}$\\
J3& 57.32 & $0.69$ & $0.24^{+0.02}_{-0.02}$ & $0.70^{+0.01}_{-0.01}$ & $0.002^{+0.29}_{-0.55}$& $0.88^{+1.12}_{-0.44}$&$0.01^{+0.03}_{-0.03}$\\
J4& 60.91 & $0.73$ & $0.24^{+0.02}_{-0.02}$ & $0.70^{+0.01}_{-0.01}$ & $-0.007^{+0.29}_{-0.55}$& $0.89^{+1.15}_{-0.45}$&$-0.01^{+0.03}_{-0.03}$\\
\multicolumn{8}{|c|}{Gaussian prior on $h=0.678\pm0.009$}\\
J1 & 58.85 & $0.70$ & $0.25^{+0.02}_{-0.02}$ & $0.67^{+0.007}_{-0.007}$ & $-0.02^{+0.33}_{-0.57}$& $0.74^{+0.92}_{-0.35}$&$-0.05^{+0.02}_{-0.02}$\\
J2 & 48.45 & $0.58$ & $0.33^{+0.04}_{-0.06}$ & $0.67^{+0.008}_{-0.008}$ & $0.26^{+0.12}_{-0.51}$& $2.10^{+2.42}_{-1.42}$&$0.05^{+0.03}_{-0.03}$\\
J3& 54.82 & $0.66$ & $0.26^{+0.03}_{-0.03}$ & $0.68^{+0.007}_{-0.007}$ & $0.09^{+0.28}_{-0.59}$& $0.91^{+1.35}_{-0.49}$&$-0.04^{+0.02}_{-0.02}$\\
J4& 55.19 & $0.66$ & $0.25^{+0.02}_{-0.03}$ & $0.67^{+0.007}_{-0.007}$ & $0.06^{+0.27}_{-0.57}$& $0.91^{+1.25}_{-0.48}$&$-0.06^{+0.02}_{-0.032}$\\
\hline
\end{tabular}}
\label{tab:mpcparjoint}
\end{table}


\begin{figure}
\centering
\includegraphics[width=7cm,scale=0.45]{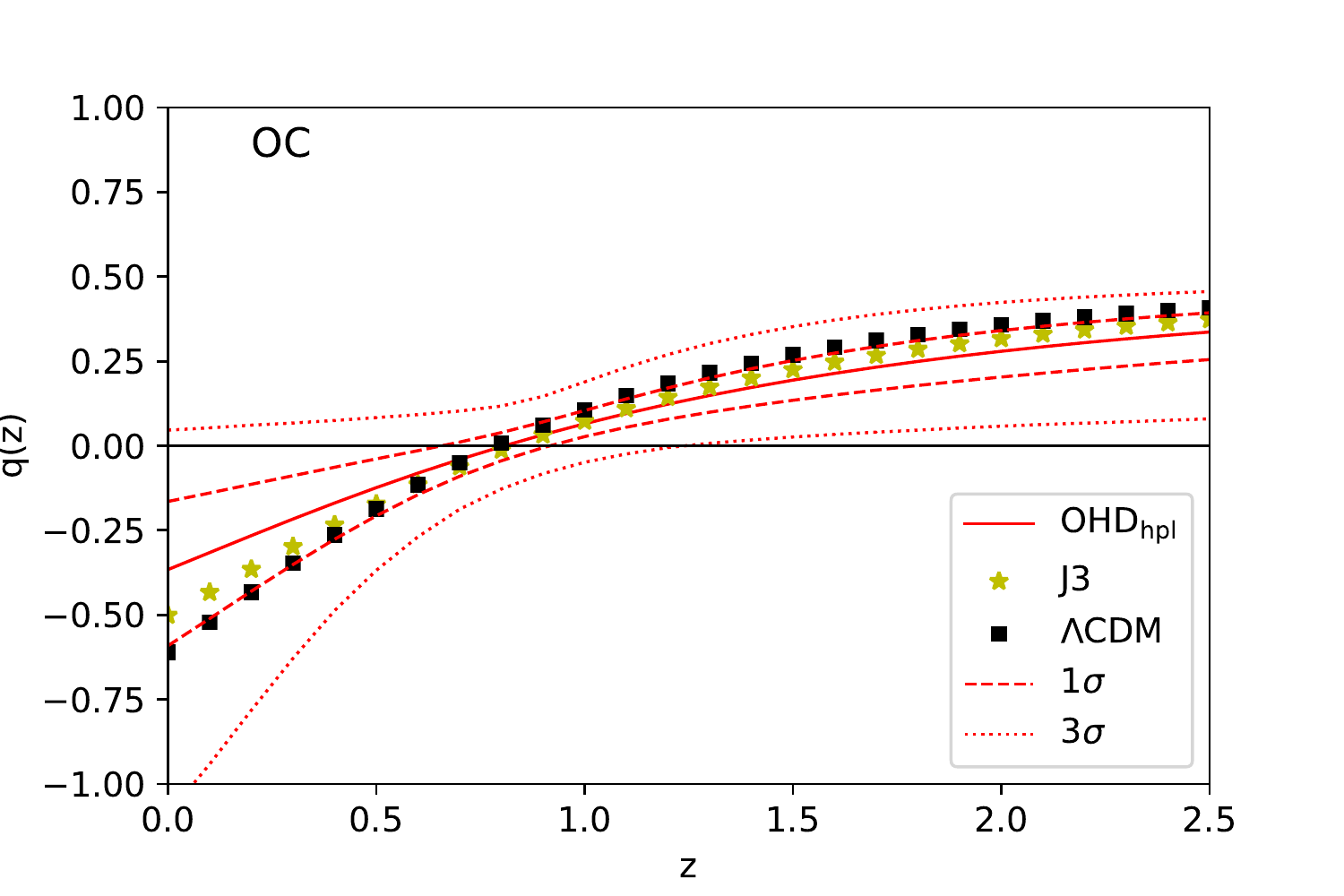} 
\includegraphics[width=7cm,scale=0.45]{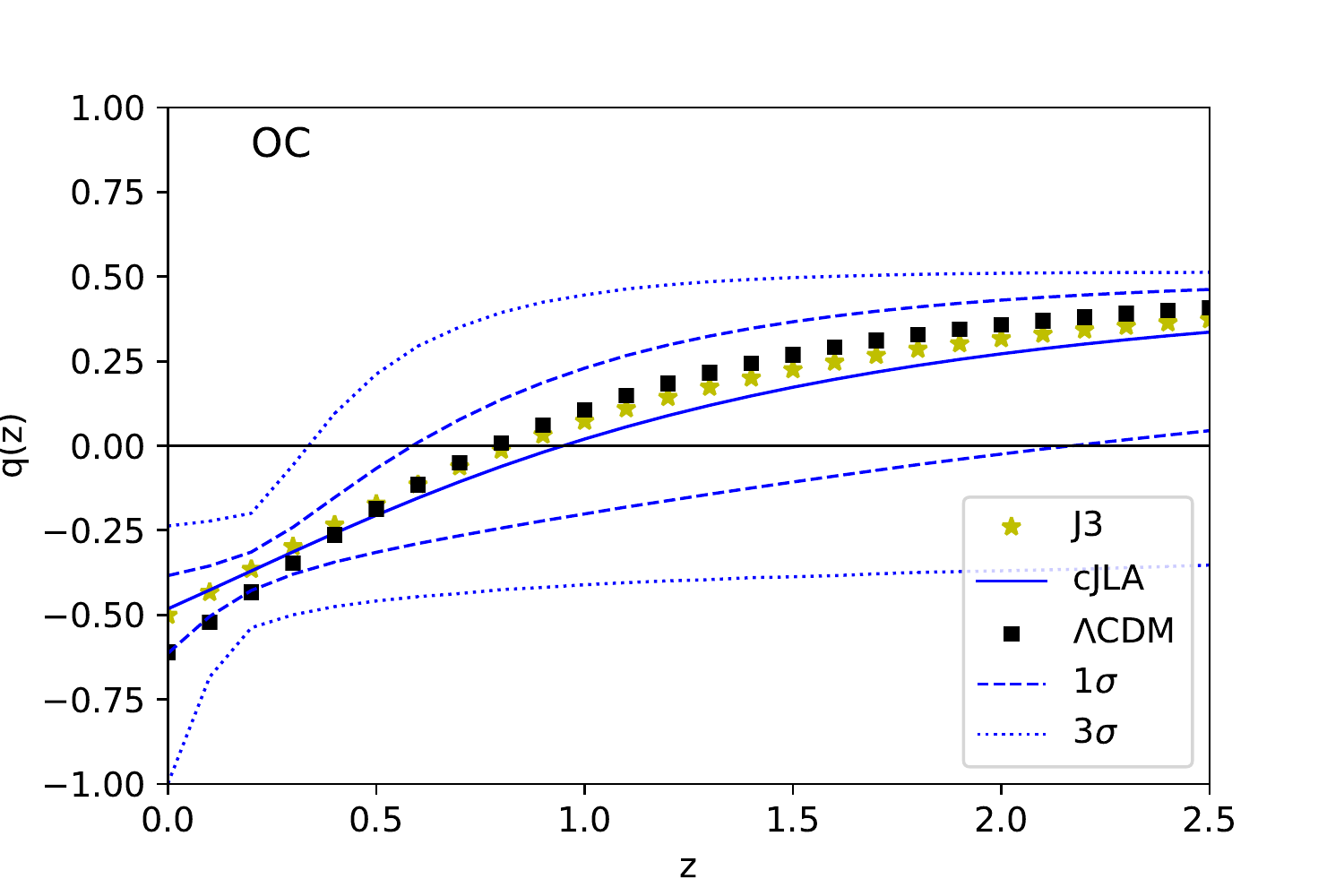} 
\caption{Reconstruction of the deceleration parameter $q(z)$ for the OC model and $\Lambda$CDM using the constraints from OHD$_{\mathrm{hpl}}$ (top panel) and cJLA data (bottom panel) with a flat prior on $h$. The $q(z)$ reconstruction from J3 constraints is shown in both panels. The dashed and the dotted lines represent the $68\%$ and $99.7\%$ confidence levels respectively.}
\label{fig:qzoc}
\end{figure}

To confirm that the OC model can drive to a late cosmic acceleration, we reconstructed the deceleration parameter using the mean values derived from the different data sets. Figure \ref{fig:qzoc} shows that the $q(z)$ dynamics is similar for the $\Lambda$CDM and OC models when the OHD$_{\mathrm{hpl}}$, cJLA and J3 constrains are used, i.e., the universe has a late phase of accelerated expansion. Notice that although the confidence levels in the $q(z)$ reconstruction obtained from the SNIa constraints are bigger that those from the OHD$_{\mathrm{hpl}}$, they are consistent. The difference could be explained by the extra free parameter (nuisance) in the SNIa analysis. 

\begin{figure*}
\centering
MPC model\par\smallskip
\begin{tabular}{cc}
\subfloat[Flat prior on $h$]{\includegraphics[width=0.5\textwidth]{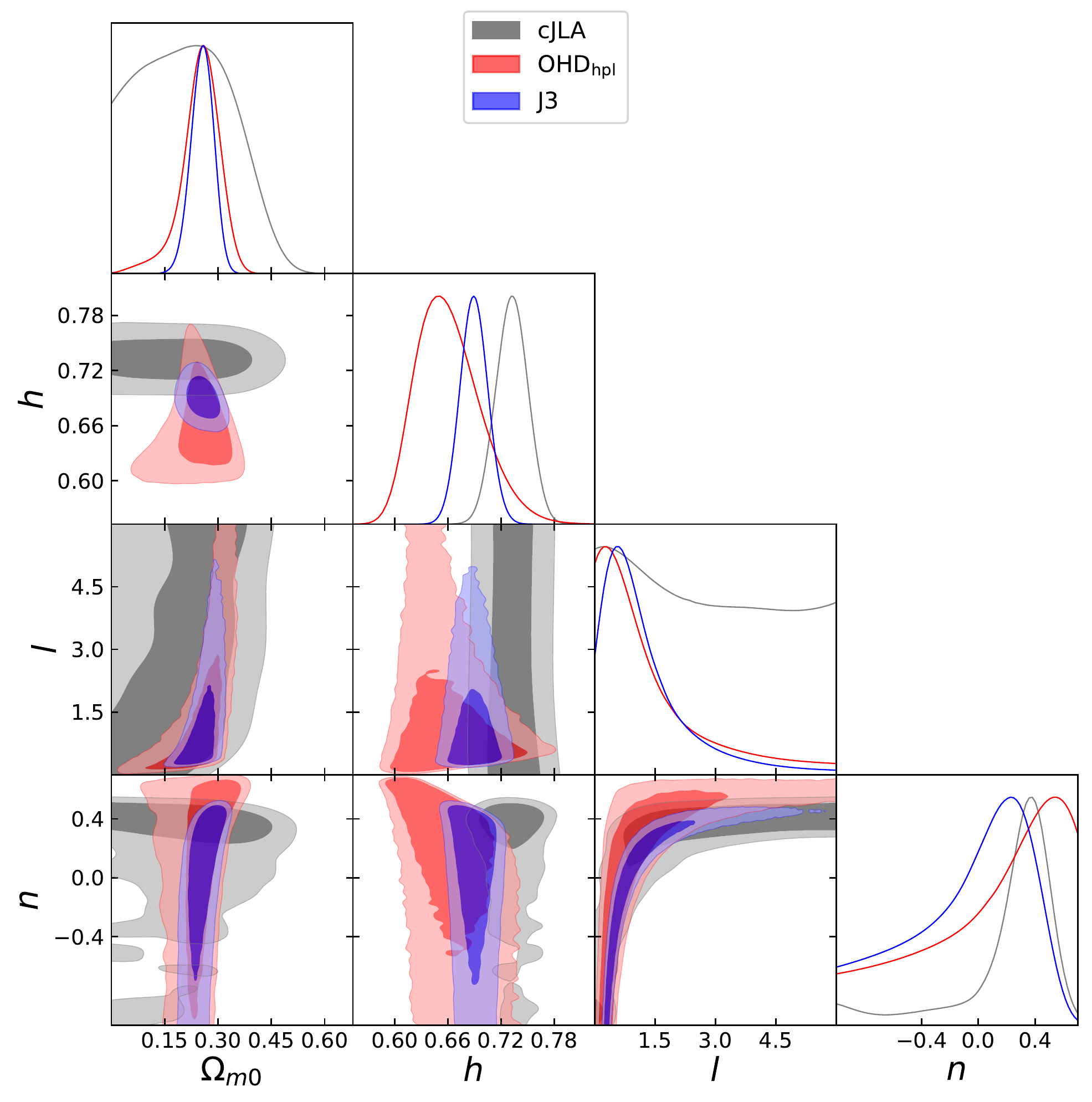}} & 
\subfloat[Gaussian prior on $h$]{\includegraphics[width=0.5\textwidth]{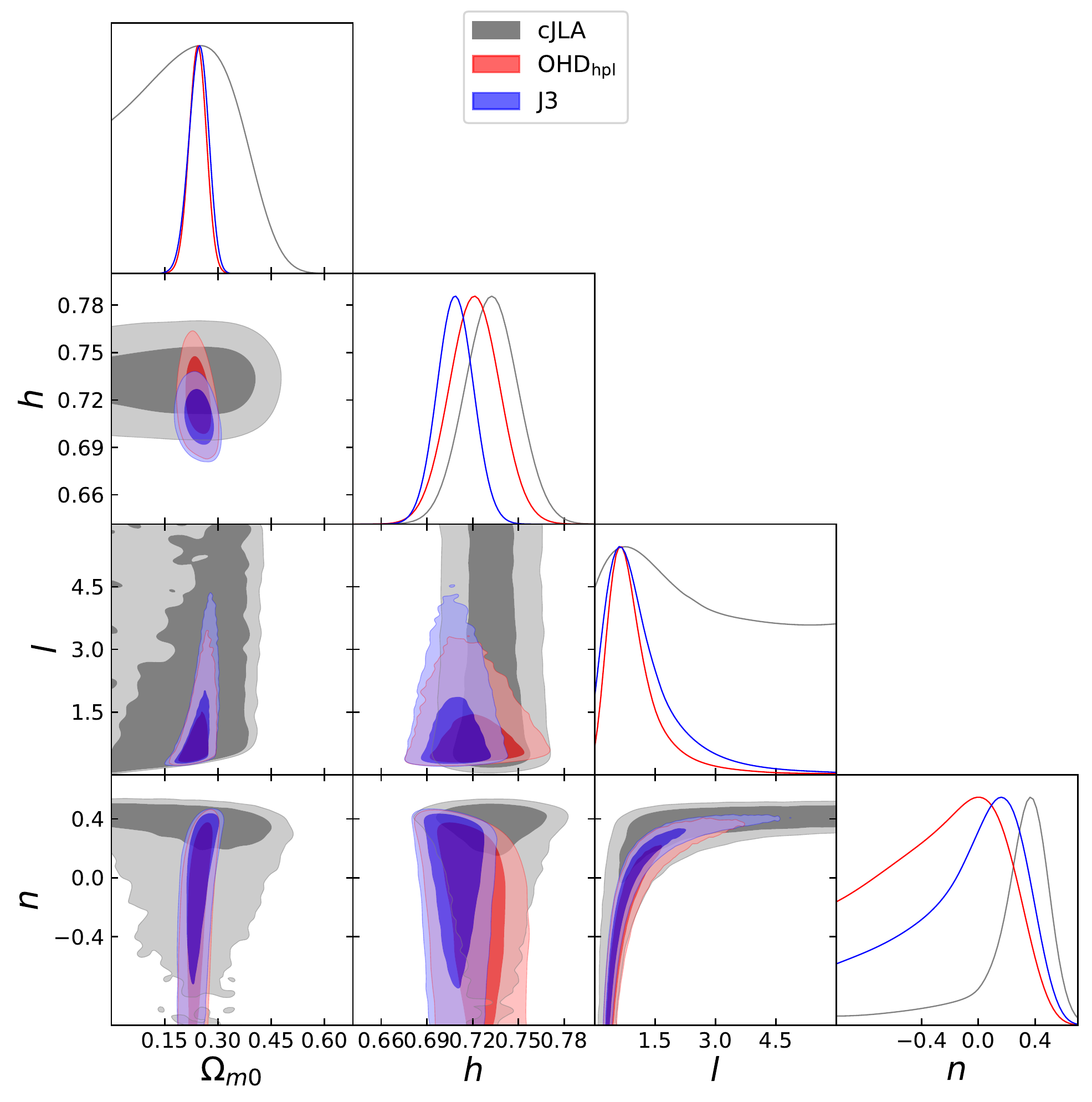}} \\
\end{tabular}
\caption{1D marginalized posterior distributions and the 2D $68\%$, $95\%$, $99.7\%$ confidence levels for the $\Omega_{m0}$, $h$, $n$, and $l$ parameters of the MPC model assuming a flat and Gaussian (h$_\mathrm{Riess}$) prior on $h$.}
\label{fig:MPCcontours}
\end{figure*}

\begin{figure}
\centering
\includegraphics[width=7cm,scale=0.45]{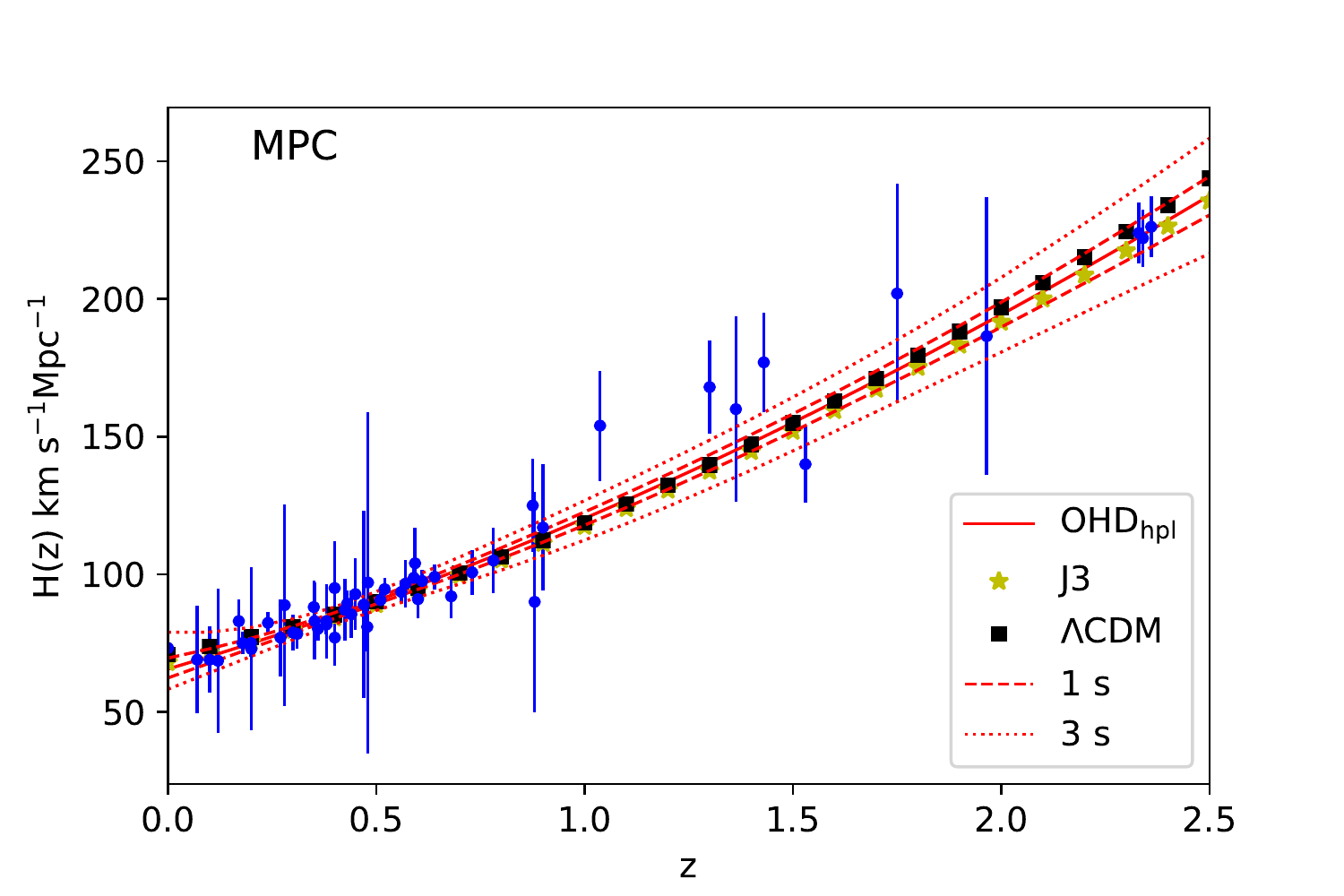} 
\includegraphics[width=7cm,scale=0.45]{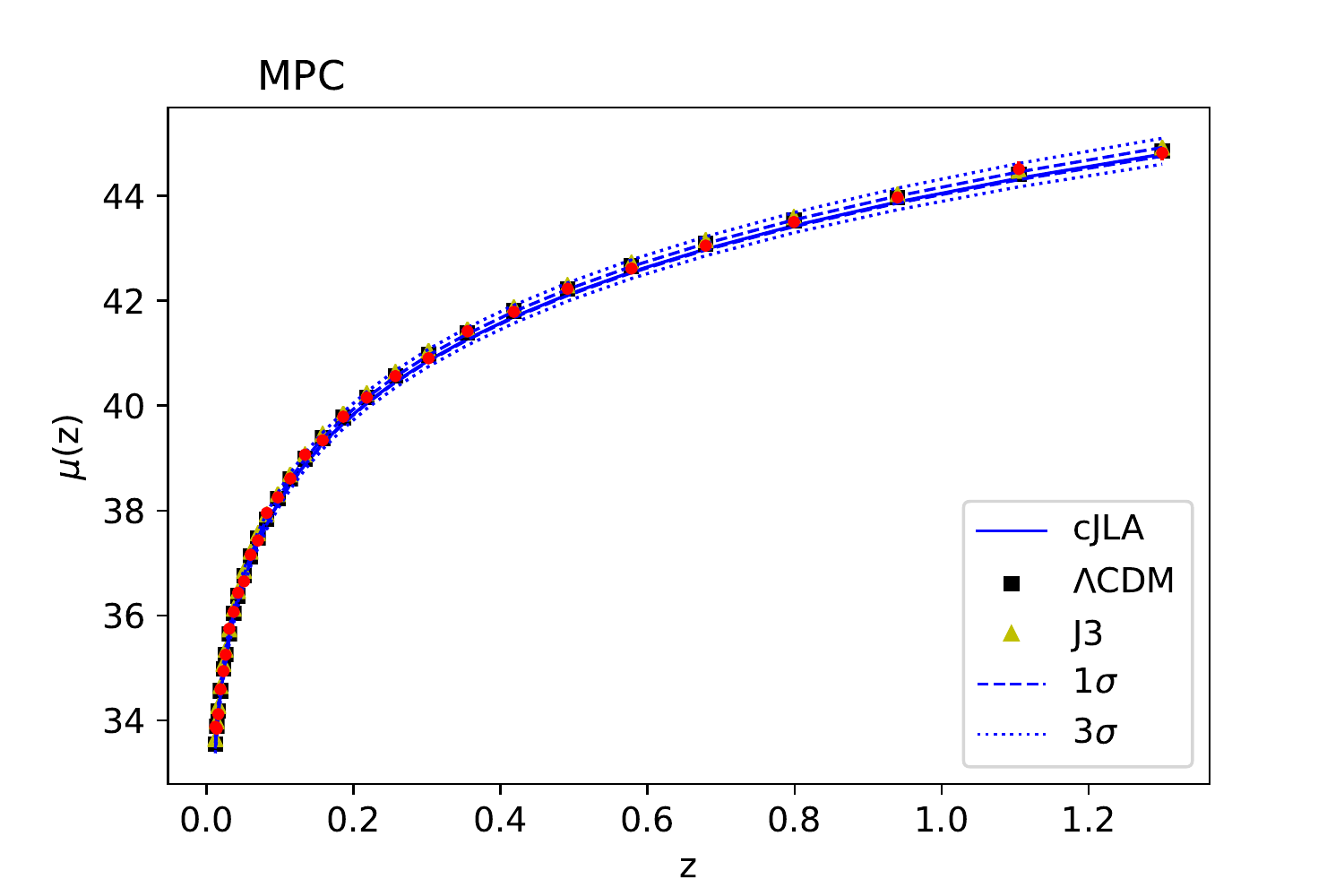} 
\caption{Fitting to OHD$_{\mathrm{hpl}}$ (top panel) and SN Ia data (bottom panel) using the mean values from the OHD$_{\mathrm{hpl}}$ (red solid lines), cJLA (blue solid lines) and J3 (yellow star and triangle) analysis for $\Lambda$CDM model (black squares) and MPC model when a flat prior on $h$ is considered.} The dashed-lines 
and dotted-lines represent the $68\%$ and $99.7\%$ confidence levels respectively. 
\label{fig:hzsnmpc}
\end{figure}

\begin{figure}
  \centering
          \includegraphics[width=7cm,scale=0.45]{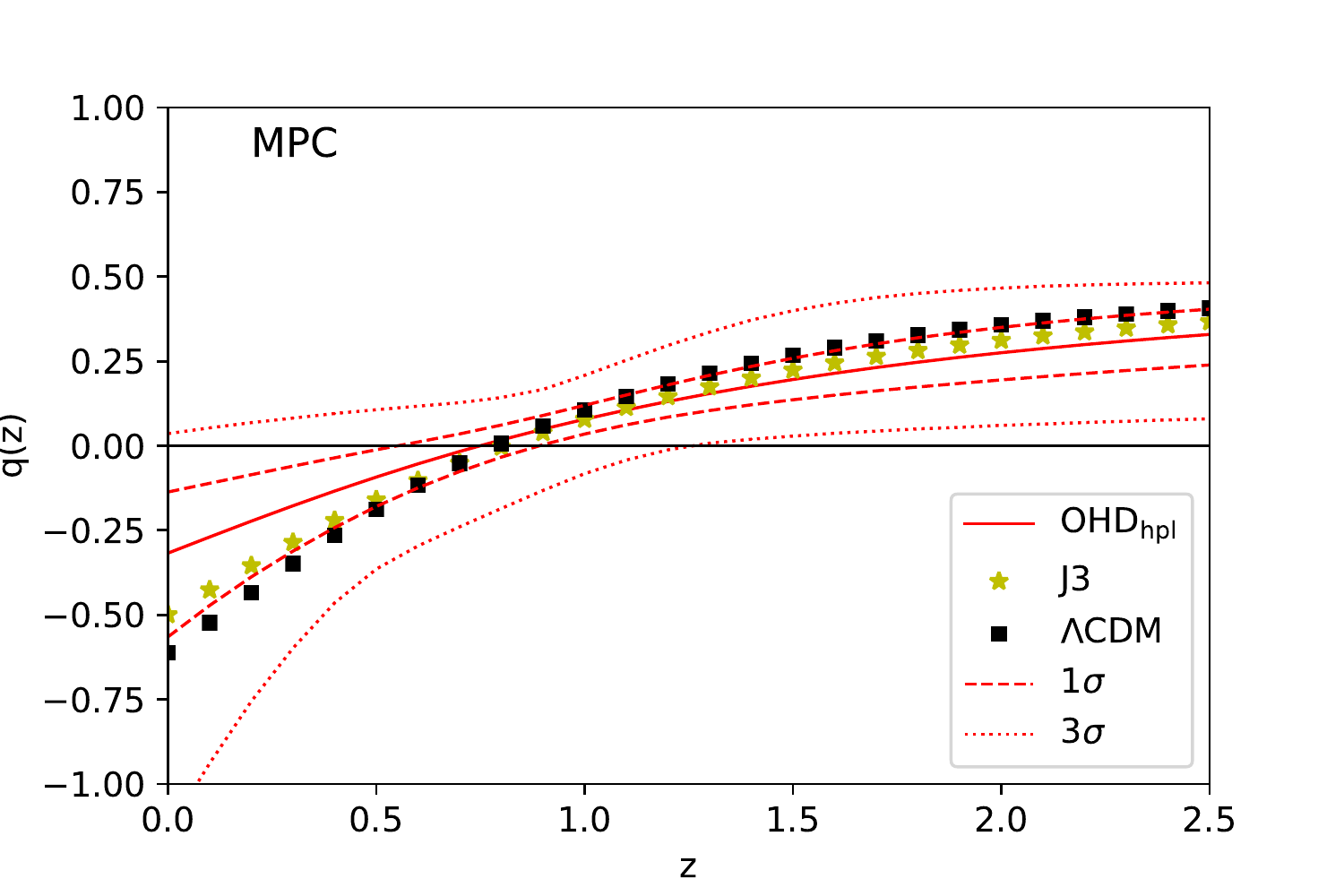} 
            \includegraphics[width=7cm,scale=0.45]{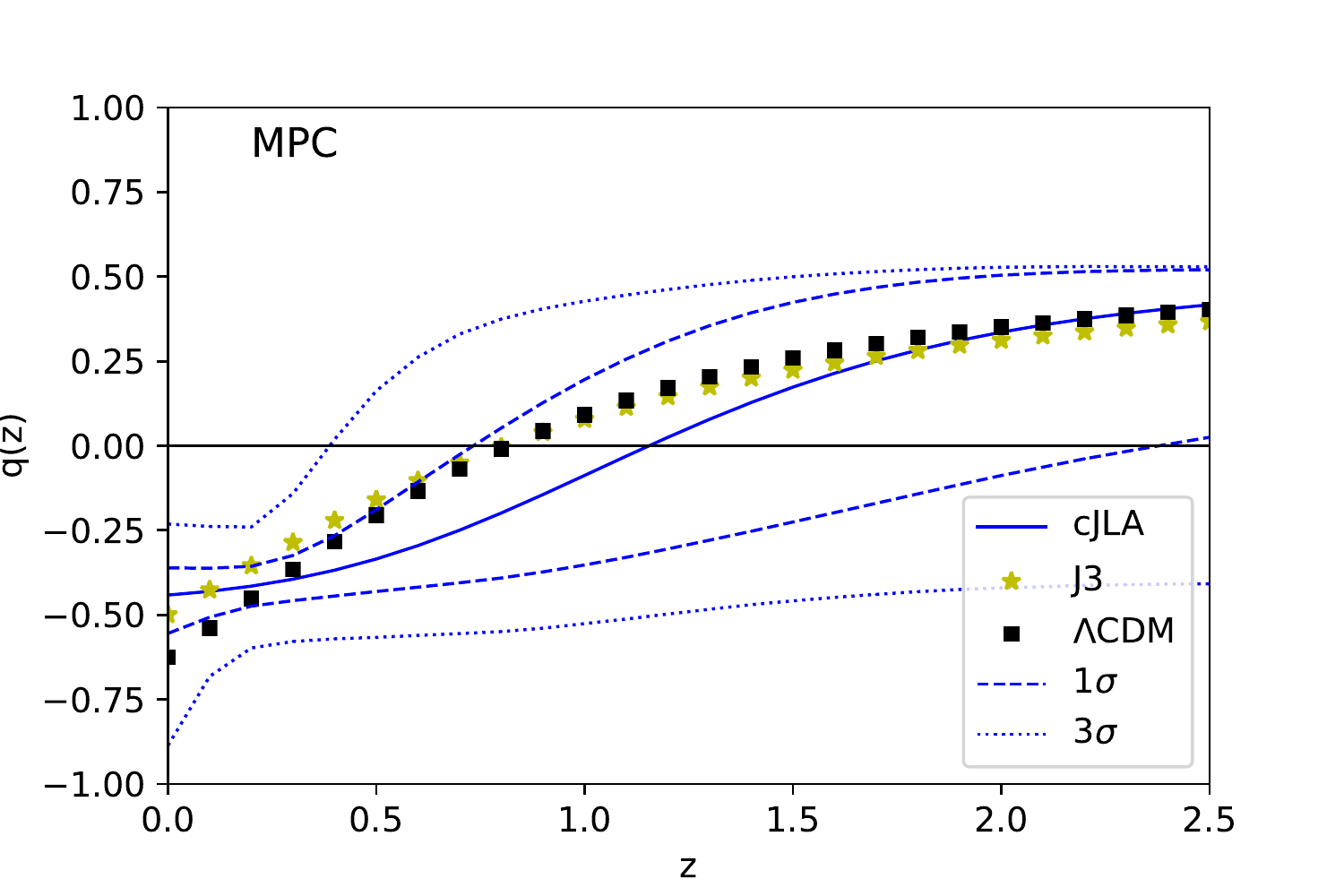} 
\caption{Reconstruction of the deceleration parameter $q(z)$ for the MPC model and $\Lambda$CDM using the constraints from OHD$_{\mathrm{hpl}}$ (top panel) and SN Ia data (bottom panel) when a flat prior on $h$ is considered. The $q(z)$ reconstruction from J3 constraints is shown in both panels. The dashed-lines and dotted-lines represent the $68\%$ and $99.7\%$ confidence levels respectively.} 
\label{fig:qzmpc}
\end{figure}

Figure \ref{fig:MPCcontours} shows the 1D marginalized posterior distributions and the 2D $68\%$, $95\%$, $99\%$ contours for the $\Omega_{m0}$, $h$, $n$ and $l$ parameters of the MPC model obtained from OHD$_{\mathrm{hpl}}$, cJLA, and J3 with flat (left panel) and Gaussian (right panel) priors on $h$. Considering a flat prior on $h$, the different data sets provide slightly different constraints on $\Omega_{m0}$ and $h$. For instance, the $\mathrm{OHD_{DA}}$ estimates higher (lower) values on $\Omega_{m0}\, (h)$ and SN Ia lower (higher) values. However, the limits are consistent within the $1\sigma$ C. L. For the $n$ and $l$ constraints, we also obtained a marginal tension using different data but they are consistent within the $1\sigma$ C. L. Notice that our constraints include $n=0$ and $l=1$, which reproduces the $\Lambda$CDM dynamics. All our bounds are similar within the $1\sigma$ C.L. to those obtained by other authors, e.g. \citet{Li:2012ApJ} combining SN Ia, BAO and CMB data measure $n=0.014^{+0.36}_{-0.94}$, $l=1.09^{+1.01}_{-0.46}$, \citet{Magana:2015A} using strong lensing features estimate $n=0.41\pm{0.25}$, $l=5.2\pm{2.25}$,
\citet{Zhai:2017} provide $n=0.16^{+0.08}_{0.09}$, $l=1.38^{+0.25}_{-0.22}$ from the joint analysis of CMB, BAO plus SN Ia (JLA) data, and \citet{Zhaiz:2017} give $n=0.02^{+0.26}_{-0.41}$, $l=1.1^{+0.8}_{-0.4}$ from the joint analysis of CMB, BAO, SN Ia, $f\sigma_{8}$ and the $H_{0}$ value from \citet{Riess:2016}. In addition, the $\chi_{red}^{2}$ values point out that the $\mathrm{OHD_{DA}}$ provides better (unbiased) MPC constraints and the values from SN Ia data suggest that their errors (cJLA sample) are underestimated. 
Considering the Gaussian prior on $h$ by \citet{Riess:2016jrr}, the OHD, OHD$_{\mathrm{hpl}}$, and OHD$_{\mathrm{hw9}}$ probes yield improvements in the MPC constraints (see the $\chi_{red}$ values). For the SN Ia (cJLA) test, there is no significant difference with the flat prior case. Notice that the stringent limits are estimated from the joint analysis (see also Fig. \ref{fig:MPCcontours}).
Figure \ref{fig:hzsnmpc} shows the fittings to the OHD$_{\mathrm{hpl}}$ and cJLA data using the OHD$_{\mathrm{hpl}}$, cJLa and J3 constraints of the MPC parameters and those of the $\Lambda$CDM model with a flat prior on $h$. To propagate the errors on OHD, $\mu(z)$, and $q(z)$, we have used a Monte Carlo approach.
For both, OHD and $\mu(z)$ fittings, there is no significant statistical difference between the MPC model and the standard one. In addition, a good agreement at $1\sigma$ is obtained employing the J1, J2 and J4 constraints. In addition, Figure \ref{fig:qzmpc} shows the reconstruction of the $q(z)$ parameter using the constraints from the OHD and SN Ia data.
For the OHD constraints, the $q(z)$ dynamics for the MPC is in agreement with that of the standard model. When the SN Ia estimations are used, the history of the cosmic acceleration for the MPC model is consistent with the $\Lambda$CDM within the $1\sigma$ and $3\sigma$ C.L. 
Thus, the MPC scenario is viable to explain the late cosmic acceleration without a dark energy component and its cosmological dynamics is almost indistinguishable from the standard model. 


\section{Conclusions and Outlooks} \label{Con}

In this paper we analyze two alternatives to explain the late cosmic acceleration without a dark energy component: the original (OC) and modified polytropic Cardassian (MPC) models which are also excellent laboratories to study deviations from GR. The Cardassian models establish the modification of the canonical Friedmann equation as a consequence of a braneworld dynamics which emerges from novel ideas of the space-time dimensions and is based on a generalized Einstein-Hilbert action. 

To constrain the exponents $n$ and the $n-l$ of the OC and MPC models, we used 51 observational Hubble data, 740 SNIa data points of the JLA sample (fJLA) and 31 binned distance modulus of the compressed JLA sample (cJLA). The OHD compilation contains $31$ points measured using the differential age technique in early-type-galaxies and $20$ points from clustering. These last points are biased due to an underlying $\Lambda$CDM cosmology to estimate the sound horizon at the drag epoch, which is used to compute $H(z)$. Moreover, these data points are estimated taking into account very conservative systematic errors. Therefore, we constructed two homogenized and model-independent samples for the clustering points using a common $r_{d}$ obtained from Planck and WMAP measurements.

We found that the different OHD samples provide consistent constraints on the OC and MPC parameters. In addition, there is no significant differences on the constraints obtained from the cJLA and those estimated from fJLA. Furthermore, we obtained consistent constraints at $3\sigma$ confidence level when different Gaussian priors on $h$ are employed. We performed a joint analysis with the combination of cJLA and one homogenized OHD sample.
Our results shown that the OC and MPC free parameters are consistent with the traditional dynamics dictated by the Friedmann equation (see Tables \ref{tab:ocparnueva}-\ref{tab:mpcparjoint}) containing a cosmological constant (CC). However, in the Cardassian models the extra terms in the canonical Friedmann equation mimic the CC but it comes from the n-term of the energy momentum tensor, unlike in the traditional form where the CC is added by hand in the Friedmann equation. Of course, those problems affecting the CC will be transferred to the interpretation of n-dimensional geometry and, as a consequence, to the emerging of the n-term of the energy-momentum tensor. Therefore, the idea is to interpret and to know the global topology of our Universe to generate a solution for the DE problem and the current Universe acceleration. 

\section*{Acknowledgments}
We thank the anonymous referee for thoughtful remarks and suggestions.
J.M. acknowledges support from CONICYT/FONDECYT 3160674.
M.H.A. acknowledges support from CONACYT PhD fellow, Consejo Zacatecano de Ciencia, Tecnolog\'{\i}a e Innovaci\'on (COZCYT) and Centro de Astrof\'{\i}sica de Valpara\'{\i}so (CAV). M.H.A. thanks the staff of the Instituto de F\'{\i}sica y Astronom\'{\i}a of the Universidad de Valpara\'{\i}so where part of this work was done.
M.A.G.-A. acknowledges support from CONACYT research fellow, Sistema Nacional de Investigadores (SNI) and Instituto Avanzado de Cosmolog\'ia (IAC) collaborations.

\bibliographystyle{mnras}
\bibliography{librero0}

\appendix
\section{Compressed JLA sample}
\begin{table}
\centering
\begin{tabular}{|cc|}
\hline
$z_{b}$&$\mu_{b}$\\
\hline
0.010 & 32.953886976\\
0.012 & 33.8790034661\\
0.014 & 33.8421407403\\
0.016 & 34.1185670426\\
0.019 & 34.5934459829\\
0.023 & 34.9390265264\\
0.026 & 35.2520963261\\
0.031 & 35.7485016537\\
0.037 & 36.0697876073\\
0.043 & 36.4345704737\\
0.051 & 36.6511105942\\
0.060 & 37.1580141133\\
0.070 & 37.4301732516\\
0.082 & 37.9566163488\\
0.097 & 38.2532540406\\
0.114 & 38.6128693372\\
0.134 & 39.0678507056\\
0.158 & 39.3414019038\\
0.186 & 39.7921436157\\
0.218 & 40.1565346033\\
0.257 & 40.5649560582\\
0.302 & 40.9052877824\\
0.355 & 41.4214174356\\
0.418 & 41.7909234574\\
0.491 & 42.2314610669\\
0.578 & 42.6170470706\\
0.679 & 43.0527314851\\
0.799 & 43.5041508283\\
0.940 & 43.9725734093\\
1.105 & 44.5140875789\\
1.300 & 44.8218674621\\
\hline
\end{tabular}
\caption{Compressed JLA sample which contains 31 binned distance modulus fitted to the full JLA sample by \citet{Betoule:2014}. The first column is the binned redshift and the second column is the binned distance modulus.}
\label{tab:jlacom}
\end{table}

\end{document}